\documentclass[twocolumn]{aastex631}

\received{January 25, 2024}
\revised{February 26, 2024}
\accepted{March 7, 2024}

\shorttitle{Optical-NIR Continuum Emission of PSR and PWN in SNR\,0540-69.3}
\shortauthors{Tenhu et al.}
\graphicspath{{./}{figures/}}

\urldef{\footurldeconv}\url{https://scikit-image.org/docs/stable/api/skimage.restoration.html#skimage.restoration.richardson_lucy}
\urldef{\footurlconv}\url{https://docs.scipy.org/doc/scipy/reference/generated/scipy.ndimage.convolve.html#scipy-ndimage-convolve}
\urldef{\footurlkernel}\url{https://photutils.readthedocs.io/en/stable/api/photutils.psf.matching.create_matching_kernel.html}

\begin{document}

\title{Spatial Variations and Breaks in the Optical-NIR spectra of the Pulsar and PWN in SNR\,0540-69.3}

\correspondingauthor{L. Tenhu}
\email{lcetenhu@kth.se}

\author[0000-0002-7746-8512]{L. Tenhu}
\affiliation{Department of Physics, KTH Royal Institute of Technology, The Oskar Klein Centre, AlbaNova, SE-106 91 Stockholm, Sweden}

\author[0000-0003-0065-2933]{J. Larsson}
\affiliation{Department of Physics, KTH Royal Institute of Technology, The Oskar Klein Centre, AlbaNova, SE-106 91 Stockholm, Sweden}

\author[0000-0003-1546-6615]{J. Sollerman}
\affiliation{Department of Astronomy, Stockholm University, The Oskar Klein Centre, AlbaNova, SE-106 91 Stockholm, Sweden}

\author[0000-0002-3664-8082]{P. Lundqvist}
\affiliation{Department of Astronomy, Stockholm University, The Oskar Klein Centre, AlbaNova, SE-106 91 Stockholm, Sweden}

\author[0000-0001-6815-4055]{J. Spyromilio}
\affiliation{European Southern Observatory, Karl-Schwarzschild-Strasse 2, D-85748 Garching, Germany}

\author[0000-0002-3464-0642]{J. D. Lyman}
\affiliation{Department of Physics, University of Warwick, Gibbet Hill Road, Coventry CV4 7AL, UK}

\author{G. Olofsson}
\affiliation{Department of Astronomy, Stockholm University, The Oskar Klein Centre, AlbaNova, SE-106 91 Stockholm, Sweden}



\begin{abstract}
\noindent The supernova remnant SNR\,0540-69.3, twin of the Crab Nebula, offers an excellent opportunity to study the continuum emission from a young pulsar and pulsar-wind nebula (PWN). We present observations taken with the VLT instruments MUSE and \mbox{X-shooter} in the wavelength range 3000–25,000\,\AA, which allow us to study spatial variations of the optical spectra, along with the first near-infrared (NIR) spectrum of the source. We model the optical spectra with a power law (PL) $F_\nu\propto\nu^{-\alpha}$ and find clear spatial variations (including a torus-jet structure) in the spectral index across the PWN. Generally, we find spectral hardening toward the outer parts, from $\alpha\sim1.1$ to $\sim0.1$, which may indicate particle reacceleration by the PWN shock at the inner edge of the ejecta or alternatively time variability of the pulsar wind. The optical-NIR spectrum of the PWN is best described by a broken PL, confirming that several breaks are needed to model the full spectral energy distribution of the PWN, suggesting the presence of more than one particle population. Finally, subtracting the PWN contribution from the pulsar spectrum we find that the spectrum is best described with a broken-PL model with a flat and a positive spectral index, in contrast to the Crab pulsar that has a negative spectral index and no break in the optical. This might imply that pulsar differences propagate to the PWN spectra.
\end{abstract}



\section{Introduction} 
\label{sec:intro}

Some remnants originating from a core-collapse supernova (SN) are observed to contain a pulsar surrounded by a pulsar-wind nebula (PWN). PWNe are formed when a pulsar wind, comprising relativistic particles and electromagnetic fields, meets the inner edge of the material ejected in the SN explosion. Young pulsars\,($\lesssim$\,10\,kyr) and their PWNe primarily emit synchrotron radiation, though inverse Compton scattering can contribute at the highest energies (e.g., \citealt{Gaensler2006,Vink_20}).

Continuum emission from pulsars and their PWNe can be characterized by power-law (PL) models relating the emitted flux to the corresponding frequencies, $F_\nu \propto \nu^{-\alpha}$, where $\alpha$ determines the spectral slope and is called the spectral index. Generally, several spectral indices are needed to model PWN continuum emission from the radio to gamma-rays (e.g., \citealt{Gaensler2006,Vink_20}). The spectral breaks and indices contain information about the conditions in pulsars and PWNe, including their particle energy and density distributions as well as inhomogeneities within the PWN \citep{Reynolds2017}. 

The ideal targets for continuum emission studies (especially in the optical range) are young and bright supernova remnants (SNRs) emitting over a broad wavelength range, located at high galactic latitudes (minimal extinction), and with resolved PWNe. The Large Magellanic Cloud (LMC) remnant SNR\,\mbox{0540-69.3} (hereafter SNR\,0540) fulfills these conditions, being \mbox{1100--1200} years old \citep{Reynolds_85,Larsson_21, Lundqvist2022} and $\sim50$\,kpc away \citep{P19}. 

SNR\,0540 has been observed from radio to X-rays (e.g., \citealt{Kirshner_1989,Manchester_93,Gotthelf2000,Williams_2008,Morse2006,Lundqvist2020,Larsson_21}) and its PWN can be resolved with current instruments (having a diameter of $\sim4''$). SNR\,0540 is also dubbed the `Crab\,twin’ \citep{Clark_82,Petre_07}. These two remnants share many similarities; they are of similar age, contain both pulsars and PWNe, with the pulsars having comparable spin periods in the millisecond range, 30 ms for the Crab  \citep{Lyne2015} and 50 ms for the SNR\,0540 pulsar \citep{Seward_1984,Marshall_16}, as well as comparable pulsar energy-loss rates. 

Despite these similarities, the differences are significant. SNR\,0540 belongs to the class of oxygen-rich SNRs, and has an outer shell associated with the forward shock, emitting in radio, mm and X-rays, located at $\sim30''$ ($\sim 7$\,pc) \citep{Manchester_93,Hwang2001,Brantseg2014,Lundqvist2020}. By contrast, no outer shell has been detected at these wavelengths for the Crab (e.g., \citealt{Frail95,Seward2006Crab,Hitomi2018}), although there is evidence for a fast shell from \ion{C}{4} $\lambda$1550 \citep{Sollerman2000}. The current understanding is that SNR\,0540 is the remnant of a Type\,II explosion with a massive progenitor ($\gtrsim15\,M_\odot$, \citealt{Chevalier2006,Williams_2008}). Observations of the Crab have shown that the amount of mass and the velocity of the filaments can only account for a low energy explosion ($\sim10^{50}$\,erg), which would point to an electron capture SN instead of a classical type II (\citealt{Yang2015} and references therein).

Searching for spectral breaks and/or spatial variations in the continuum emission requires a broad wavelength range and observations that enable the removal of line emission contribution. Previous studies of the PWN in SNR\,0540 (hereafter PWN\,0540) in the infrared (IR) and optical wavelengths (e.g., \citealt{Serafimovich_2004,Williams_2008,Mignani2012}) have used imaging and long-slit spectroscopy, which have restricted the spectroscopic data availability to narrow wavelength ranges (imaging) or small spatial regions (long-slit spectroscopy). Generally, limited spectral and spatial resolution make it challenging to accurately remove line contamination or pulsar contribution, respectively, from the PWN continuum spectrum. 

Similarly, several efforts have been made to measure the optical continuum spectral index for the pulsar in SNR\,0540 (PSR B0540-69, hereafter PSR\,0540, e.g., \citealt{Middleditch1987,Hill_97,Serafimovich_2004,Serafimovich2005,Mignani_2010,Mignani2012,Mignani2019}) but contamination by emission lines and challenges in removing the PWN contribution have prevented forming a consensus. Additionally, there have not been any spectroscopic studies of SNR\,0540 in the near infrared (NIR). All these difficulties have yielded conflicting results for the shapes of the PSR and PWN\,0540 optical continuum spectra and how they are connected to the multiwavelength spectra at shorter and longer wavelengths.

It is also interesting to study how the PWN continuum spectrum varies spatially, as this provides information about particle transport and energy losses. Spectral index mapping has been done for several PWNe, primarily in X-rays (e.g., \citealt{Mori_2004,Guest2020,Hu2022}). In the case of PWN\,0540, \citet{Petre_07} and \citet{Lundqvist2011} mapped the spectral index in X-rays and reported the canonical spatial softening of the continuum spectrum toward the outer regions of the nebula, attributed to synchrotron losses. On the other hand, there are few studies that map the continuum slope of PWNe at optical wavelengths (e.g., \citealt{Veron-Cetty1993}). Using photometry, \citet{Serafimovich_2004} studied the spatial variations in the optical continuum spectrum of PWN\,0540. They focused on eight different regions within the PWN and found evidence for possible spatial hardening toward the PWN\,0540 outer boundary. A more detailed spatial mapping of the continuum spectrum would help understand the underlying particle acceleration environments in the PWN and shed light on how the optical emission from different regions connects to the emission at shorter wavelengths.

PSR\,0540 was observed to experience a spin-down rate change in late 2011 \citep{Marshall_15}. This event makes it interesting to compare observations from pre and post-2011 epochs due to possible changes in the properties of PSR and PWN\,0540. As an example, \citet{Ge_19} observed significant brightening of the PWN\,0540 flux in X-rays after the 2011 event. Therefore, multiwavelength post-2011 observations are important in understanding the current state of PSR and PWN\,0540.

In this study we present and discuss observations of PWN\,0540 and PSR\,0540 in SNR\,0540 covering the wavelength range from blue (UVB) to NIR (\mbox{3000--25,000}\,Å). We use data obtained in 2019 from the \mbox{X-shooter} spectrograph and MUSE integral-field spectrograph, which are both mounted on the Very Large Telescope (VLT) of the European Southern Observatory (ESO) in Chile. These data provide the first NIR spectrum of this source along with the possibility to study spatial variations of the spectra in the optical wavelengths. This paper, being the second in a series of SNR\,0540 studies with \mbox{X-shooter} and MUSE data, continues the work of \citet{Larsson_21} which we hereafter refer to as L21. 

We organize the paper as follows. In Section~\ref{sec:obs} we describe the observations and data reduction procedure. Section~\ref{sec:construction_of_the_continuum_spectra} discusses the process of constructing the continuum spectrum. In Section~\ref{sec:results}, we present the results and analysis and follow up with a discussion in Section~\ref{sec:discussion}. Finally, we summarize and conclude the paper in Section~\ref{sec:conclusions}. A comparison of the MUSE and \mbox{X-shooter} spectra and a study of systematic effects is provided in Appendix \ref{app:muse_vs_xshooter}, while Appendix \ref{app:extinction} presents additional details on the extinction toward SNR\,0540, and Appendix \ref{app:pulsar_age} focuses on the birth spin period of the pulsar.

\section{Observations and data reduction} 
\label{sec:obs}

We analyze observations of SNR\,0540 obtained with the Multi Unit Spectroscopic Explorer (MUSE, \citealt{MUSE}) and \mbox{X-shooter} \citep{X-shooter} at the VLT. Details about the observations and data processing are provided in L21. Here we summarize the most important points relevant for our analysis. 

The MUSE observations were performed in January and March 2019 and provided a total exposure time~of~$\sim$\,2.35\,h. The resulting spectra cover the wavelength range \mbox{4650--9300\,\AA} (with a gap between \mbox{5760--6010\,\AA} caused by the Na laser), with a spectral resolution $R=$\,\mbox{1750--3750}. The wide-field mode adaptive optics (WFM-AO) was used for these observations, providing a field of view of 1\farcm{0}$\times$1\farcm{0} sampled at 0\farcs{2} per spatial pixel (spaxel). The image quality averaged over wavelength as measured in the MUSE frame is 0\farcs{8} (FWHM).

\begin{figure}[t!]
\plotone{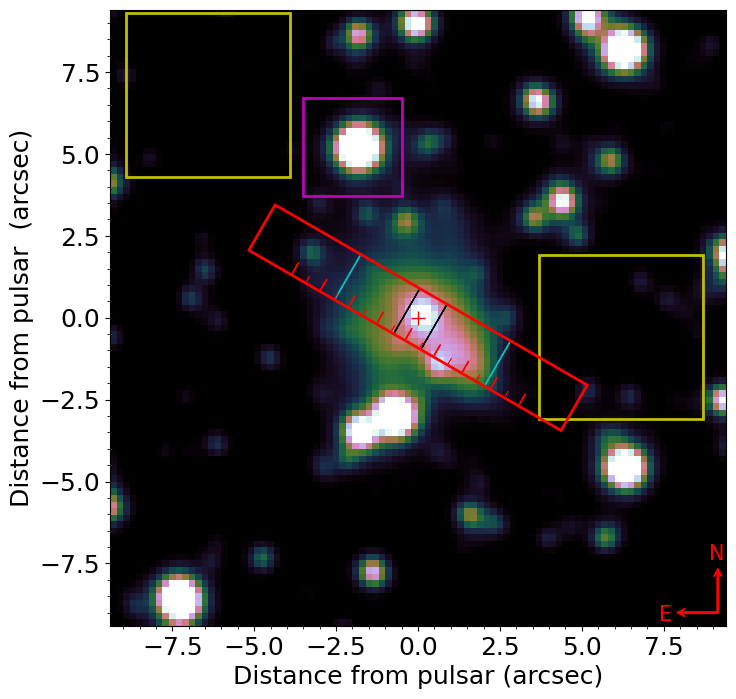}
\caption{MUSE image of SNR\,0540 continuum emission (color code denoting the intensity). The pulsar (red cross) is located at the center. The red rectangle shows the \mbox{X-shooter} slit (width = 1\farcs{6}, UVB band,  position angle $60^\circ$). The major tick marks along the slit label 1\farcs{0} distance and the minor tick marks 0\farcs{5}. The black lines label the region that is used to study the pulsar continuum emission (width 0\farcs{96}, extending 0\farcs{4} to the NE and 0\farcs{56} to the SW from the pulsar) and the cyan lines show the nebula region (width $5\farcs{28}$, extending to $2\farcs{48}$ to the NE and $2\farcs{80}$ to the SW from the pulsar). The magenta rectangle ($3\farcs{0}\times3\farcs{0}$) defines the PSF region and the yellow rectangles (both $5\farcs{0}\times5\farcs{0}$) the background regions.}
\label{fig:slit}
\end{figure}

\mbox{X-shooter} is a spectrometer with three arms, in the UVB, visible (VIS), and NIR bands, which together cover the wavelength range \mbox{3000--25,000}\,Å. The \mbox{X-shooter} observations were performed between 29 October and 2 November 2019 with total exposure times of 6720, 4560, and 7200\,s in the UVB, VIS, and NIR, respectively, and  with slit widths $1\farcs{6}$ (UVB), $1\farcs{5}$ (VIS), and $1\farcs{2}$ (NIR). This set-up provides a spectral resolution of 3200, 5000, and 4300, together with pixel sizes of 0\farcs{16}, 0\farcs{16}, and 0\farcs{21} for each arm, respectively. The slit dimensions compared to the source and its orientation are shown superposed on the MUSE continuum image in \autoref{fig:slit}. During the observations, the airmass was 1.5 and the seeing 0\farcs{8}--1\farcs0. The resulting data were reduced in the STARE mode.

\section{Construction of the Continuum Spectrum}
\label{sec:construction_of_the_continuum_spectra}

This section focuses on constructing the continuum spectrum so that its shape can be investigated. In Section \ref{subsec:psf}, we describe the convolution applied to account  for the MUSE Point Spred Function (PSF) wavelength dependence, as well as the deconvolution applied to improve the spatial resolution. Next, in Section \ref{subsec:spectral-lines}, we introduce the method to remove all emission lines in order to construct the continuum spectrum and follow by presenting the extinction correction for all spectra in Section \ref{subsec:LMC_extinction}.

\subsection{PSF Wavelength-dependence corrections for MUSE} 
\label{subsec:psf}

The width of the MUSE PSF varies with wavelength from FWHM  \mbox{$\sim1\farcs{0}$ at $4770$\,\AA} to FWHM $0\farcs{7}$ at $7992$\,\AA,\footnote{MUSE data longward of 8000\,Å were excluded, see Section \ref{subsec:spectral-lines}.} though we note that the MUSE PSF is highly non-gaussian and the FWHM is therefore just a rough estimate. This PSF variation causes a hard (i.e. low spectral index) halo around point sources and therefore has to be accounted for when studying the spatial variations of the spectral index. We choose to convolve all 2D slices in the MUSE data cube (each 2D slice corresponding to a wavelength) so that all PSFs match with the widest PSF width. The \mbox{X-shooter} PSF is not observed to vary significantly with wavelength suggesting it is dominated by seeing and the instrumental resolution. Thus we do not convolve the X-shooter spectra in this work. Below, we describe the convolution/deconvolution processes performed for the MUSE data.

We choose a star, $\sim5''$ north from the pulsar (\autoref{fig:slit}) at wavelength $4770$\,\AA, and with a sampling interval of $15\times15$ pixels ($3\farcs{0}\times3\farcs{0}$), as the target (widest and bluest) PSF for the convolution process. We subtract the background emission by defining two background regions, $25\times25$ pixel ($5\farcs{0}\times5\farcs{0}$) squares, in the northwest (NW) and east (E, \autoref{fig:slit}). For each 2D slice, we subtract the background with the median flux value of these two background regions.

To accomplish the PSF matching required for convolution, we define a matching kernel (with a `Hanning' window) to stretch the narrower PSFs at all wavelengths to the bluest and widest PSF with the help of the python package photutils.psf.\footnote{\footurlkernel} These matched kernels are then provided for the convolution algorithm as the target PSFs. The convolution is performed with the python package scipy.ndimage\footnote{\footurlconv} with convolution mode `nearest'. We verify the performance of the algorithm by comparing images and inspecting the radial profiles of a sample of field stars at different wavelengths after the convolution. 

\begin{figure*}[t!]
\plotone{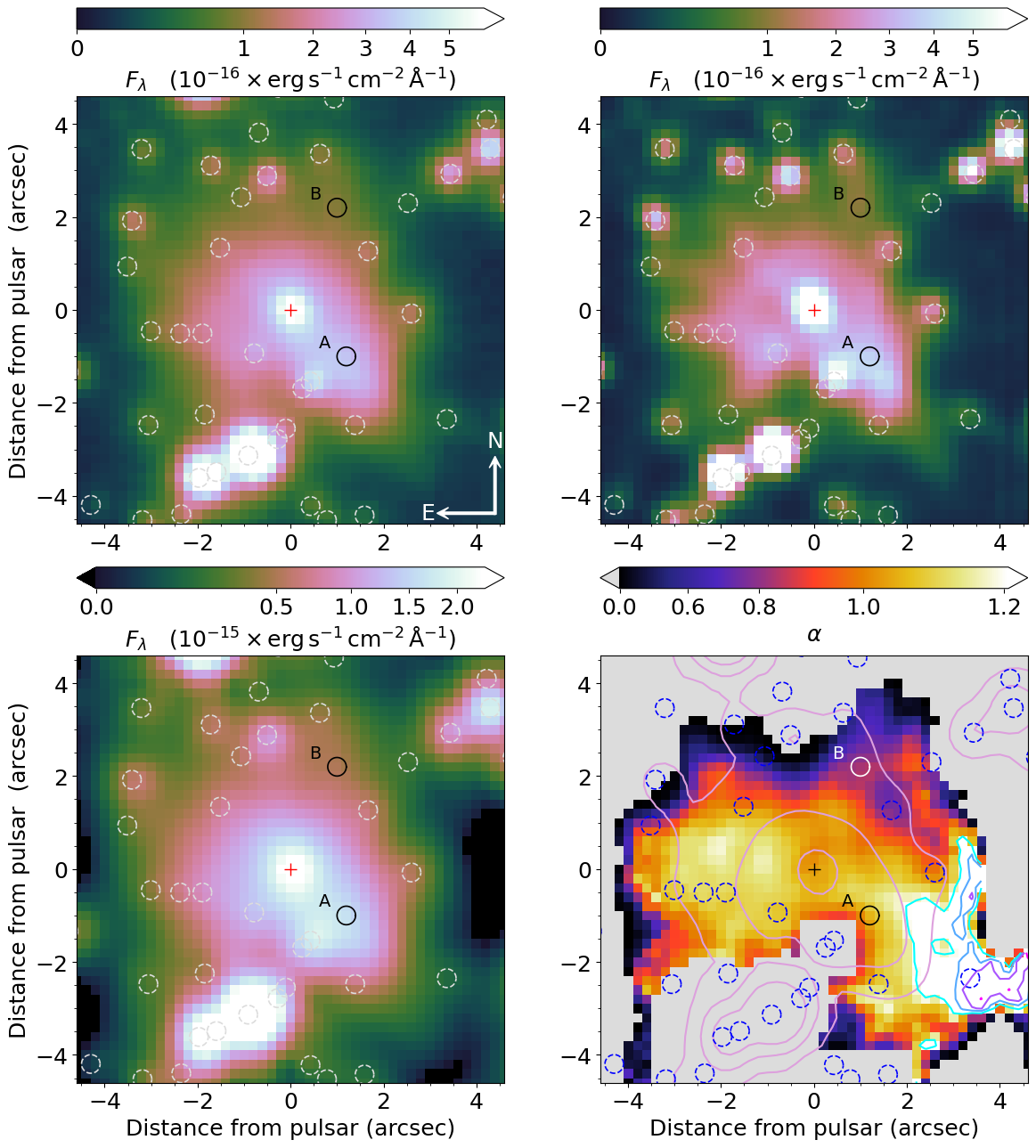}
\caption{\textit{Top left:} MUSE continuum flux image produced by integrating over the reddest quarter of the continuum wavelength range (7236--7992\,Å), where the spatial resolution is the best. \textit{Top right:} The MUSE continuum fluxes from the top left panel deconvolved. \textit{Bottom left:} The whole range of MUSE continuum fluxes (4770--7992\,Å) convolved to match the lowest spatial resolution. The color scales of all the images have a square-root base. \textit{Bottom right:} Map of spectral indices, $\alpha$, obtained by fitting the convolved continuum fluxes from the bottom left panel by a PL model. The color scale has a power-law base with power~$=2.5$. The contours from cyan to magenta correspond to spectral index values ${\alpha = 1.2, 1.4, 1.6}$, and $1.8$, respectively. Grey areas are masked due to bright field stars, spectral index $\alpha<0$ or spectral index uncertainty $\sigma_\alpha>0.2$ and are excluded from the subsequent analyses. Flux levels from the bottom left panel are superposed as pink contours for comparison. In all panels, the pulsar is located at the origin and marked with a red or black cross. The dashed circles indicate locations where field stars have been identified and these locations are excluded in the subsequent analysis. The cross (pulsar position) and the solid circles denoted by A and B show the spaxels from which indicative fits to the data, shown in \autoref{fig:muse_example_fit}, are extracted. We note that all spectral analysis is performed on the convolved cube only.
\label{fig:pixel_map}}
\end{figure*}

The convolved image of the PWN~0540 continuum emission is shown in the bottom left panel of \autoref{fig:pixel_map}, in comparison with the original non-convolved image in the top left panel of the same figure (summed over only the reddest quarter of the continuum emission wavelengths, which provides the best spatial resolution). All subsequent analyses of the MUSE observations are performed with the convolved data unless otherwise mentioned.

Both the wavelength dependence of the MUSE PSF, and the convolution process to correct for said dependence will hide many details related to the spatial morphology of the SNR. Therefore, we also produce a deconvolved image of the MUSE emission. As opposed to the convolution process, deconvolution of the 2D slices requires a target PSF that has the best spatial resolution, i.e. the reddest and narrowest PSF. We thus choose the same star (and sampling interval) as in the convolution process, but at wavelength $7992$\,\AA. We use this PSF to deconvole each slice of the reddest quarter of the continuum cube (\mbox{7236--7992 Å}), noting that the PSF does not change significantly over this wavelength interval.

We apply the Richardson-Lucy (RL) deconvolution method (\citealt{Richardson-1972,Lucy-1974}, for recent usage in astronomy see \citealt{Sakai-2023} and references therein) by using the algorithm included in the python package scikit-image.\footnote{\footurldeconv} This deconvolution algorithm performs best with non-negative input values; hence we remove all the non-physical negative flux values from the 2D slices before deconvolution. We allow the deconvolution process to run for 15 iterations so that possible enhancment of noise can be avoided. As a result, we provide the highest spatial resolution MUSE image of the continuum emission of PWN~0540 in the top right panel of \autoref{fig:pixel_map}.

The improved spatial resolution acquired with deconvolution comes with a noteworthy caveat: it is unclear how much noise is enhanced in the process. We note, though, that deconvolution uncertainties do not propagate into our quantitative analysis. All measurements are performed on the convolved data and the deconvolved image is used solely as an aid in identifying the locations of emission components within the PWN.

\subsection{Isolation of the Continuum Spectrum} 
\label{subsec:spectral-lines}

\begin{figure*}[t!]
\plotone{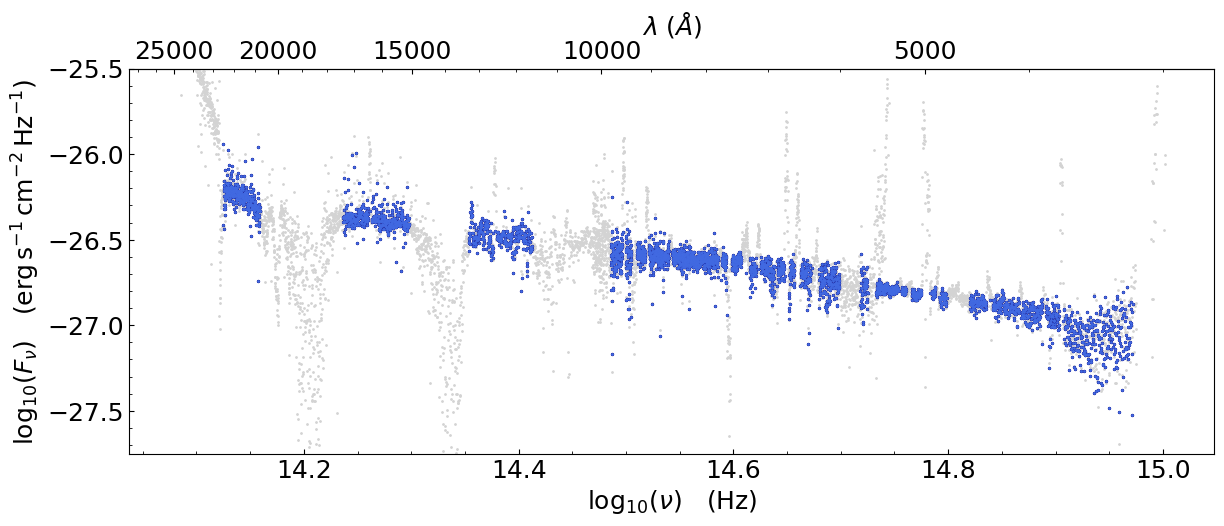}
\caption{\mbox{X-shooter} UVB-NIR continuum emission of the PWN. The extraction region is defined in \autoref{fig:slit}. Blue points show the resulting continuum spectrum after the masking procedure (see text for details). Grey points are masked data.}
\label{fig:xshooter_continuum}
\end{figure*}

\begin{figure*}[t!]
\plotone{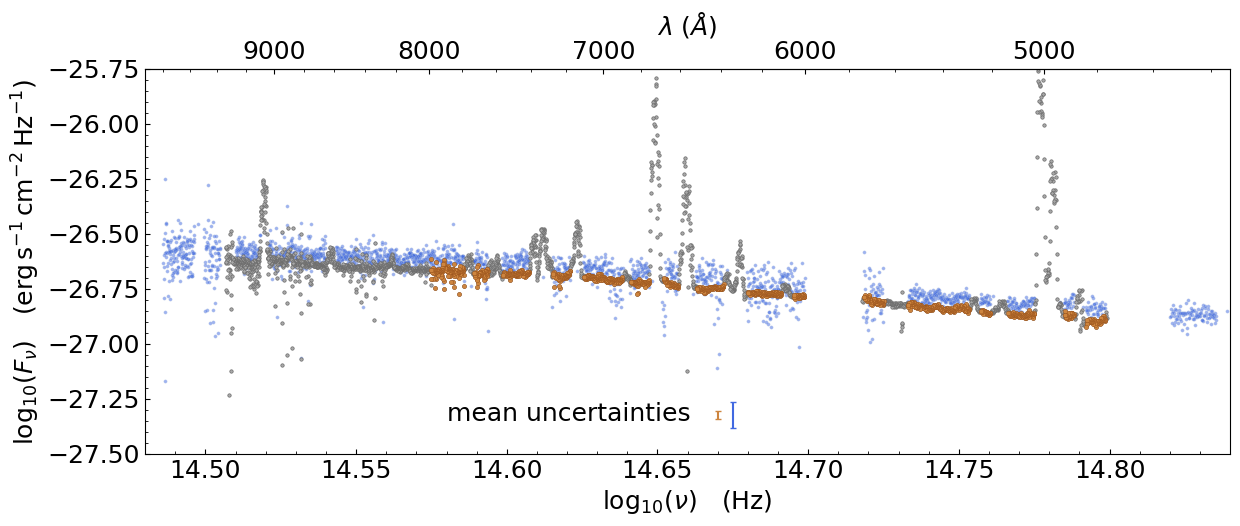}
\caption{MUSE PWN continuum emission (golden points) extracted through an \mbox{X-shooter} pseudo-slit, shown together with the \mbox{X-shooter} continuum spectrum from \autoref{fig:xshooter_continuum} (blue). The gray points denote the excluded MUSE data. Mean flux uncertainties for both instruments are shown at the bottom.
\label{fig:muse_continuum}}
\end{figure*}

We correct all spectra for the local systemic velocity by fitting the emission lines from the interstellar medium (ISM) near the center of the remnant. The resulting velocity of $277$\,km\,s$^{-1}$ is identical for both \mbox{X-shooter} and MUSE, with standard deviations of $1.5$\,km\,s$^{-1}$ and $6.5$\,km\,s$^{-1}$, respectively. 

For \mbox{X-shooter}, we follow \citet{Sollerman2019} and remove regions of poor atmospheric transmission, after which the remaining wavelength intervals are \mbox{3200--5550}, \mbox{5650--9800}, \mbox{11,600--13,300}, \mbox{15,100--17,400}, and \mbox{20,500--22,500}\,\AA. In addition, we filter out individual pixels with extremely low or high values (more than $5\sigma$ away from the median of the continuum emission fluxes in each spectrum) for both \mbox{X-shooter} and MUSE. 

To prepare both data sets for continuum emission studies, we also remove narrow ISM/CSM (circumstellar medium) lines as well as broad ejecta lines from all spectra. These lines are removed from all the spectra by manually determining masks by studying the emission line widths, since we want to avoid assuming any particular shape for the continuum spectra (a requirement for sigma-clipping algorithms). For most of the lines, a mask with a width from $-800$ to $+1300$\,km\,s$^{-1}$ is sufficient, but for broader emission lines like [\ion{O}{3}]\,${\lambda\lambda4959,\,5007}$, we use mask widths up to \mbox{$-1700$--$+1700$\,km\,s$^{-1}$}. 

In \autoref{fig:xshooter_continuum}, we illustrate the processes of isolating the continuum spectrum and show the whole UVB-NIR \mbox{\mbox{X-shooter}} PWN continuum spectrum extracted from the PWN region shown by cyan lines (and excluding the pulsar) in \autoref{fig:slit}. Throughout this paper, we bin the \mbox{\mbox{X-shooter}} spectra by a factor of 8 for visual clarity. However, this binning is only applied when plotting, while all fits are performed on the unbinned data.

We scale the \mbox{X-shooter} fluxes from each band by the slit width ratios to compensate for the narrower slit widths in the longer wavelength ranges. This scaling process is performed by using the UVB slit width ($1\farcs{6}$) as the reference slit. However, the linear scaling based on the relative sizes of the slit widths is a first approximation since it assumes an idealised case of an extended uniform surface brightness object. We also note that this scaling is only performed for the PWN and not for the pulsar, which is discussed separately in Appendix~\ref{app:muse_vs_xshooter}.

The results of isolating the continuum in the MUSE data are shown in \autoref{fig:muse_continuum}, where the remaining wavelength range is \mbox{4800--8000\,Å}. In addition to the aforementioned masks in the relevant wavelength range for MUSE, we also mask the wavelength range \mbox{8000--9000}\,Å. This region has been reported to suffer from light contamination,\footnote{In ESO Phase 3 Data Release Description for MUSE: \url{http://www.eso.org/rm/api/v1/public/releaseDescriptions/78} it is reported that the contamination happened between 1 February and 18 April 2019. Our observations are taken between January and March 2019 and the data possess a similar bump.}  which is seen as a clear bump in our spectra. The bump is not apparent in the background-subtracted spectra in \autoref{fig:muse_continuum}, but the effect is known to vary across the field of view, so we ignore this wavelength region to avoid introducing systematic uncertainties. The MUSE spectrum in this figure is extracted by using a pseudo-slit corresponding to the \mbox{X-shooter} slit, see \autoref{fig:slit}. From \autoref{fig:muse_continuum}, one can see that the two data sets agree reasonably well.

We investigate the effects of systematics in Appendix~\ref{app:muse_vs_xshooter} by comparing how much the continuum slope differs between the two instruments. Our estimate for the magnitude of the systematic offset in the continuum slope between MUSE and \mbox{X-shooter} is $0.1$. This is likely due to a combination of differences in spatial resolution, pixel sizes, seeing, spectral resolution, as well as calibration uncertainties.

We also estimate continuum flux uncertainties in both instruments by computing standard deviations of the flux. We assume that the continuum flux does not exhibit curvature on short frequency ranges and therefore we can compute the standard deviation for short blocks of continuum emission. For this, we utilize the blocks created by the emission line mask.

\begin{figure*}[t!]
\plotone{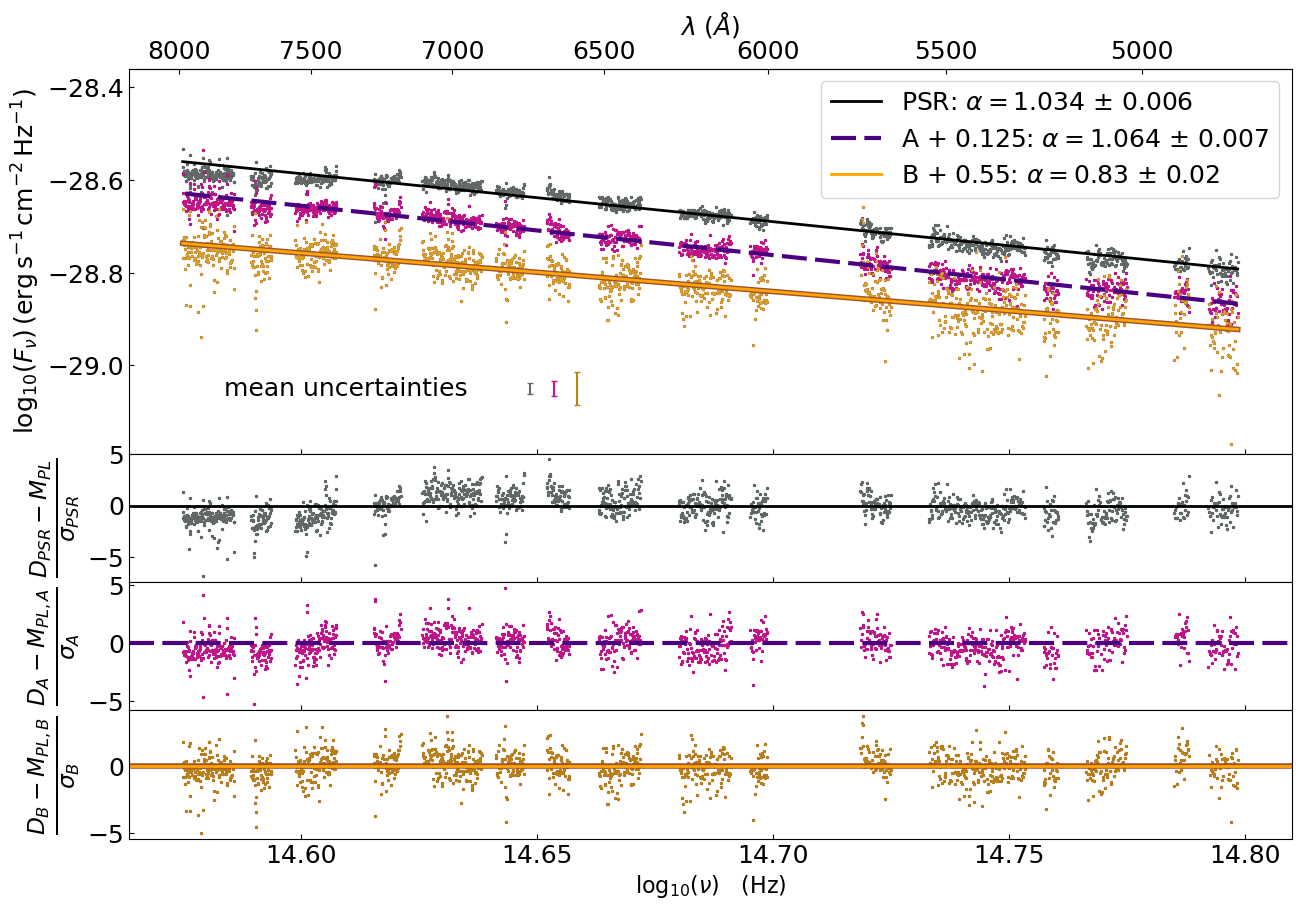}
\caption{\textit{Top panel:} MUSE continuum spectra with PL fits and fit residuals from three regions, pulsar (gray points, black solid line), blob (A, violet points, purple dashed line) and jet (B, light brown points, light brown solid line), identified in \autoref{fig:pixel_map}. Mean uncertainties for the three spectra are shown at the top center of the figure. The fit parameter uncertainties are standard deviations computed from the covariance matrix. Fluxes from A and B are lower compared to the pulsar flux and are here presented with offsets $\log_{10}\left(F_\nu\right)=$ 0.125 and 0.55\,\,erg\,s$^{-1}$\,cm$^{-2}$\,Hz$^{-1}$, respectively, for easier inspection. \textit{Three bottom panels:} PL fit residuals that are computed by subtracting the PL fit values ($M_{PL},\,M_{PL,A}$, and $M_{PL,B}$) from the corresponding measured fluxes from the PSR, A and B spaxels ($D_{PSR},\,D_{A}$, and $D_{B}$), which are then divided by the flux uncertainties ($\sigma_{PSR},\,\sigma_{A}$, and $\sigma_{B}$), respectively, for each indicative fit.}
\label{fig:muse_example_fit}
\end{figure*}

\subsection{Extinction} 
\label{subsec:LMC_extinction}

We use the analytic formula from \citet{Cardelli_1989} to correct the spectra for extinction. We use ${R_V=3.1}$ and ${E(B-V)=0.27\pm0.07}$\,mag based on an investigation of the Balmer decrement of the ISM emission in Appendix \ref{app:extinction}. This value is larger compared to the values ${E(B-V)=0.19}$\,mag or ${E(B-V)=0.20}$\,mag reported and used in previous studies, e.g. \citet{Kirshner_1989,Serafimovich_2004}. We note that there are caveats in the color excess measurement due to the complexity of the region, see Appendix~\ref{app:extinction} for discussion.

Additionally, we also study the [\ion{Fe}{2}] emission lines, 1.257\,$\mu$m and 1.644\,$\mu$m of the SNR, observed with \mbox{X-shooter} (Appendix~\ref{app:extinction}). The color excess from the iron line ratio is less certain but aligns with the result derived from the Balmer decrement.

This larger color excess value boosts especially the bluer (i.e. higher frequency) part of the spectrum and will therefore affect the spectral index results by flattening the spectrum. We study the magnitude of this effect on the MUSE wavelength range and find that the approximate difference in the spectral index is $-0.2$, when we use ${E(B-V)=0.27}$\,mag.

\section{Analysis and Results} 
\label{sec:results}

We begin by presenting the analysis of the PWN\,0540 continuum emission by first focusing on the MUSE observations, followed by the corresponding analysis with the \mbox{X-shooter} observations in Section~\ref{subsec:nebula}. Additionally, we study how the properties of the continuum spectra correlate with previous results on the line emission (L21) and polarization \citep{Lundqvist2011} in the nebula area. The continuum emission of the pulsar is analysed in Section~\ref{subsec:pulsar}.

All spectral fits are performed with least-squares and Markov Chain Monte Carlo (MCMC) routines for MUSE and \mbox{X-shooter}, respectively. We choose to use the MCMC method for \mbox{X-shooter} spectra (as is also done in \citealt{Sollerman2019}), because this method allows us to securely identify the global minimum. For the MUSE spectra, however, the least-squares method turned out to be sufficient.

Each MCMC fit is initiated with 100 samplers, which each sample for 2000 steps. After visually inspecting the MCMC chains, we discard the first 50 steps as a burn-in phase. The uncertainties on fit parameters are calculated from the fit covariance matrices (MUSE) and as $68\%$ posterior quantiles by excluding the top and bottom $16\%$ of the posterior samples (\mbox{X-shooter}) and are used as an estimate of $1\sigma$ uncertainty.

\subsection{Continuum Spectra of the PWN}
\label{subsec:nebula}

\subsubsection{MUSE}
\label{subsubsec:muse}

The top left panel of \autoref{fig:pixel_map} shows an image obtained by summing over the reddest quarter of the continuum wavelength range \mbox{(7236--7992\,Å}, extracted as described in Section~\ref{sec:construction_of_the_continuum_spectra}). We use a Hubble Space Telescope (HST) image (F547M from 2005, shown in \citealt{DeLuca_07}) as guidance to identify and exclude the field stars in the following analysis.  The continuum flux image in the top left panel of \autoref{fig:pixel_map} shows a radial decrease of flux levels, with the bright pulsar being in the center. This image also shows clear signs of a more luminous region in the southwest (SW, region around A), which has been identified as a hotspot or blob in several earlier studies (e.g., \citealt{DeLuca_07,Lundqvist2011}). In our analysis, this region will be referred to as the blob region.  Additionally, a jet-like feature (e.g., \citealt{Gotthelf2000}) is thought to be located in the NW, though we cannot confirm this from the continuum flux image (\autoref{fig:pixel_map} top left panel, region around B). However, we call this the jet region.

As described in Section~\ref{sec:construction_of_the_continuum_spectra}, we also deconvolved the reddest quarter of the MUSE continuum emission. The result is shown in the top right panel of \autoref{fig:pixel_map}. Generally, the deconvolved PWN appears more asymmetric than the corresponding non-convolved image (top left panel of \autoref{fig:pixel_map}). As expected, the deconvolution process results in more prominent point sources (pulsar and field stars) and reveals possible substructures in the inner parts of the PWN. The most significant substructre being the arc-like feature in the E. Similarly, the blob region in the SW appears more accentuated after the deconvolution.

Deconvolving a diffuse source like PWN\,0540 is uncertain, because deconvolution algorithms are predominantly highlighting point sources. This behaviour may lead to artificial features in the final image. Preventing these artefacts propagating to the subsequent analyses and results, we manage the MUSE PSF wavelength dependence by convolution (as described in Section~\ref{sec:construction_of_the_continuum_spectra}). An image of the entire convolved MUSE continuum range is presented in the bottom left panel of \autoref{fig:pixel_map}. Despite the decreased spatial resolution due to convolution, we can still identify all the features that are  present in the non-convolved image (top left panel of \autoref{fig:pixel_map}). Additionally, the areas dominated by background emission are more apparent in the convolved image due to the background subtraction performed before the convolution process. As mentioned in Section~\ref{sec:construction_of_the_continuum_spectra}, we use the convolved MUSE data for all subsequent spectral analyses.

As an integral-field spectrograph, every spaxel in the MUSE data contains a spectrum. We are thus able to treat every spaxel separately and get a continuum spectrum for each spaxel. After constructing the continuum spectra, we fit them with a single PL model ${F_\nu \propto \nu^{-\alpha}}$, where $\alpha$ is the spectral index. 

In \autoref{fig:muse_example_fit}, we present example fits (to demonstrate the quality of the fits) from three different spaxels, which we refer to as pulsar, A (located in the blob region, bottom left panel of \autoref{fig:pixel_map}) and B (located in the jet region). The example fits show that the continuum emission from each of these spaxels from different regions of the PWN is well-described by a single PL in the MUSE wavelength range. We thus proceed to fit all the spaxels in the nebula with the single PL model. We note that the example fit from the pulsar spaxel shows hints of a broken PL (bPL), and we investigate this further in Section~\ref{subsec:pulsar}, where we study the entire pulsar region instead of a single spaxel.

As can be seen in the bottom right panel of \autoref{fig:pixel_map}, the spectral index has clear spatial variation across the remnant. We mask regions where $\alpha<0$, its uncertainty $\sigma_\alpha>0.2$ and where there are bright field stars. As a general trend, the spectral index is high (i.e. soft) in the central region, $\alpha \sim 1.1$, and decreases (i.e. hardens) toward the edges of the central nebula, down to ${\alpha \sim 0.1}$. This means that, in general, the spectral index hardens radially while going toward the outer regions of the central part of the remnant.

We note that in the S, close to the blob region (and spaxel A), some residual emission of the bright field star remains after the masking (bottom right panel of \autoref{fig:pixel_map}). The spectral index of spaxel A in \autoref{fig:muse_example_fit} might therefore be slightly harder than the “pure” PWN emission at that location (by ${\sim0.04}$), though we note that the spectrum is still well described by the  PL model.

We detect several structures that have different spectral indices, for instance, an axis spanning from the northeast (NE) to the SW (including the blob region) can be identified by having a nearly constant \mbox{spectral index}\,\mbox{${\alpha\sim}$1.1--1.2}. This, in turn, could be an indication of a part belonging to a torus-like structure, which is also visible as a bright region in the flux image, especially in the SW (\autoref{fig:pixel_map}, bottom left panel). We refer to this structure as the torus hereafter. Previous studies have found evidence for the torus in optical and X-rays \citep{Morse2003,Gotthelf2000}. 

In the E at distances $\sim$1\farcs{0} and $\sim$2\farcs{0} from the pulsar, we identify two regions with slightly higher spectral index ($\alpha\sim1.2$) overlapping with the torus. In addition, to the W from the pulsar, a larger region of significantly higher spectral index of \mbox{${\alpha\sim}$1.2--1.8} can be observed. This region is overlapping with a field star in the SW but otherwise seems to belong to the PWN.

Additionally, the region in the NW (surrounding B) with \mbox{${\alpha\sim}$0.6--0.8} could be a feature of a jet, as suggested by e.g., \citet{Gotthelf2000} and \citet{Petre_07}. This kind of a structure is not apparent in the flux levels in the bottom left panel of \autoref{fig:pixel_map} as mentioned above. Also, the pulsar, the brightest point source within the central regions of the PWN, is not apparent in the spectral index map.

\begin{figure}[t!]
\plotone{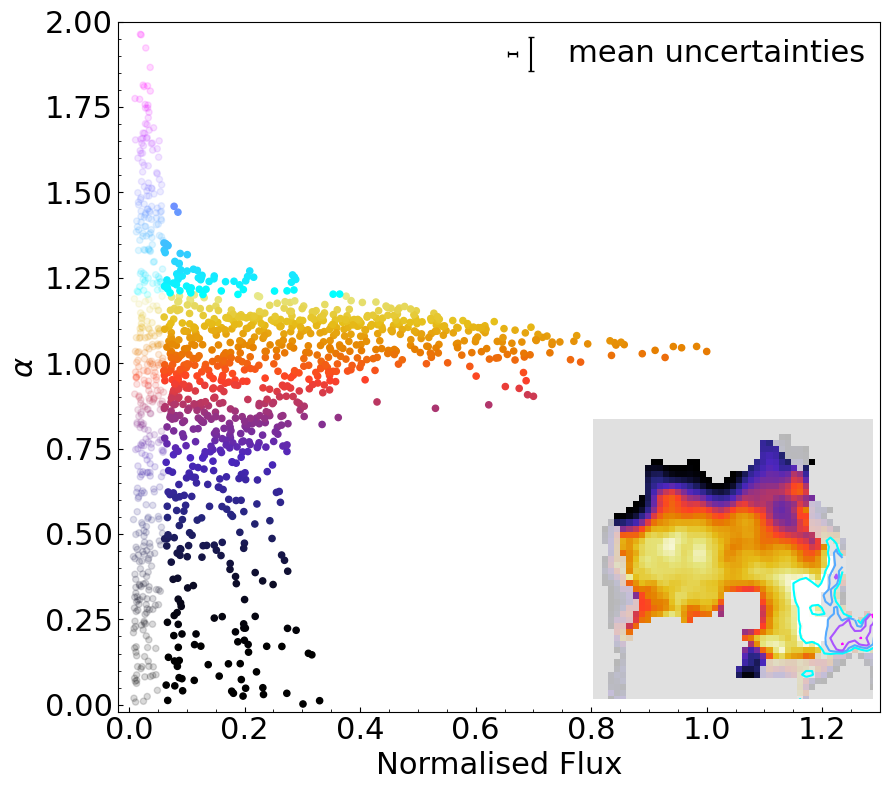}
\caption{Spectral index $\alpha$ (from \autoref{fig:pixel_map}, bottom right panel) versus convolved and normalized MUSE fluxes of PWN\,0540 (from \autoref{fig:pixel_map}, bottom left panel). Mean uncertainties for the spectral indices and fluxes (flux uncertainties multiplied by a factor of 5 for clarity) are at the top right corner. The color coding corresponds to the colors in the bottom right panel of \autoref{fig:pixel_map}, also shown in the inset. The transparent points correspond to regions where the normalised flux is less than 6\% of the maximum flux.}
\label{fig:alpha_flux}
\end{figure}

We study the relation between the continuum flux and the spectral index in \autoref{fig:alpha_flux}. The pulsar region, a 3$\times$3 spaxel square around the pulsar with ${\alpha\sim1.0}$ and normalized flux ${>0.8}$, can be distinguished from the other regions by its high flux. As expected for a point source, the spectral index in the pulsar region remains constant.

The torus, the region that in flux images is mainly apparent in the SW (overlapping A) but continues also ${\sim1\farcs{0}}$ to the NE, has an approximately constant spectral index ($\alpha\sim1.1$), while the normalized flux values vary (\mbox{$\sim0.1$--$0.8$}). This results in a plateau-like feature in the \mbox{$\alpha$--normalised-flux} space as can be seen in \autoref{fig:alpha_flux}.

Considering the spaxels with $1.2<\alpha<1.35$ in the high-$\alpha$-region reveals that this subregion most likely belongs to the torus. This connection can be made by studying the plateau-like shape in \autoref{fig:alpha_flux}, where the top of the plateau is smoothly formed by spaxels belonging to the $1.2<\alpha<1.35$ subregion. The rest of the spaxels in the high $\alpha$-region in the W and SW, where $\alpha>1.35$, behave like a background region; low flux values ($\lesssim0.1$) with highly variable spectral index.

At these low flux levels, \autoref{fig:alpha_flux} shows a wide range of continuum slopes, ranging between $\alpha\sim0$ to $\alpha\sim2$. We interpret these spaxels, also located at the edges of the PWN, to suffer from systematic effects and thus consider them more uncertain, even though the formal fit uncertainties are $\sigma_\alpha<0.2$. Finally, for the rest of the PWN, the spectral index and flux values seem to correlate when going radially outward from the edges of the torus to the the edges of the PWN.

\subsubsection{Comparison to Line Maps and Polarization}
\label{subsec:linemaps_pol}

\begin{figure*}[t!]
\plotone{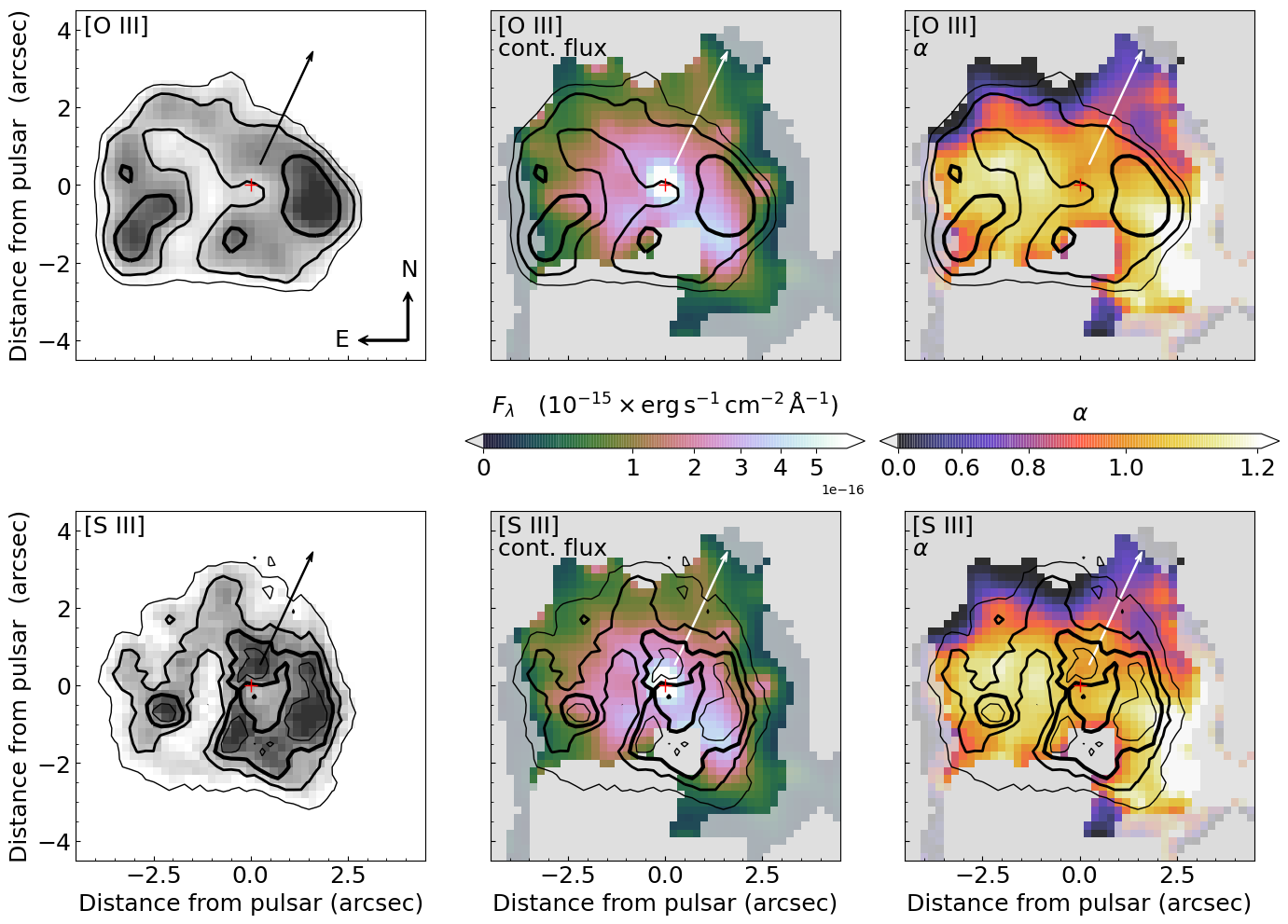}
\caption{MUSE image of [\ion{O}{3}] $\lambda$5007 emission line (\textit{top left}), [\ion{O}{3}] $\lambda$5007 emission line contours superposed on deconvolved MUSE continuum flux map (\textit{top middle}), and superposed on MUSE spectral index map (\mbox{\textit{top right}}). MUSE image of [\ion{S}{3}]\,$\lambda9069$ emission line (\mbox{\textit{bottom left}}), [\ion{S}{3}]\,$\lambda9069$ emission line contours superposed on deconvolved MUSE continuum flux map (\textit{bottom middle}), and superposed on MUSE spectral index map (\mbox{\textit{bottom right}}). The continuum emission and spectral index maps are from \autoref{fig:pixel_map}, where regions with more transparent color scale are classified as more uncertain (see \autoref{fig:alpha_flux} and text). The images of [\ion{O}{3}] $\lambda$5007 and [\ion{S}{3}]\,$\lambda9069$ emission lines are from L21. The contour levels for the line maps are set to highlight the morphology and are therefore different for the two lines. The red cross shows the pulsar's location and the arrows indicate a possible jet direction estimated from the average cavity/hole angle of the 3D emission line maps.}
\label{fig:linemap_vs_continuum}
\end{figure*}

L21 present 3D reconstructions of several emission lines from the ejecta in SNR\,0540. Here we compare the line emission from [\ion{O}{3}]\,$\lambda$5007 and [\ion{S}{3}]\,$\lambda9069$ with our results for the (deconvolved) continuum emission, presented in \autoref{fig:linemap_vs_continuum}. Overall, the continuum emission is stronger than the line emission in the center of the PWN. However, there is also significant overlap in some regions, especially on the western side, which may be caused by projection effects. 

The [\ion{O}{3}]\,$\lambda$5007 emission is more extended in the \mbox{E--W} direction than the [\ion{S}{3}]\,$\lambda9069$ emission. This is clear when compared to the continuum emission (middle panels of \autoref{fig:linemap_vs_continuum}), for instance, the two [\ion{O}{3}]-emitting blobs in the E are located close to the edges of the continuum emitting region (as seen along the line-of-sight). In the case of [\ion{S}{3}]\,$\lambda9069$, the brightest part of the corresponding blob seems to have a more central location when compared to the continuum emitting region (along the line-of-sight). In the W, the brightest continuum emission (the blob region) and the brightest emission from both lines are roughly overlapping. This agrees with the results in \citet{Lundqvist2011}, where the emission of the [\ion{S}{2}] lines is mainly concentrated to the SW overlapping with the continuum emission in that region (the [\ion{S}{2}]\,$\lambda\lambda$6716,6731 and [\ion{S}{3}]\,$\lambda9069$ are very similar, see L21).

We also estimate a possible jet direction, $\sim25^\circ$ to the W from the N and inclined by $\sim10^\circ$ away from the observer, according to the NW cavity/hole angle of the 3D emission line maps. Also \citet{Lundqvist2022} (in their Figure 6) report similar evidence for the jet direction by following the cavities of the [\ion{O}{3}] $\lambda$5007 emission line provided by \citet{Sandin2013}. The cavities of the emission lines in the SE (L21), in turn, seem to be approximately antiparallel to the jet in the NW. In addition to the emission line cavities in the possible jet (and counter jet) direction, both emission lines have less emission at the center, in the NE, and significantly low emission farther in the SW.

The emission line maps can also be compared to the spatial variation of the spectral index (\autoref{fig:linemap_vs_continuum}, right panels). Noticeable is that the brightest line-emitting regions do not exactly overlap with the hardest (low $\alpha$) continuum emission. In the SW, the brightest part of the [\ion{O}{3}] emission is located in the softest (high $\alpha$) continuum emission belonging to the torus. Also, the ring-like structure that is seen in the SW of the [\ion{S}{2}] emission is not apparent in the spectral index map, possibly due to a field star blocking the line-of-sight view. Interestingly, the cavities occurring in the NE in the line emission maps seem to be filled with a spur of softer spectral index of the NE arc. 

\begin{figure}[t!]
\plotone{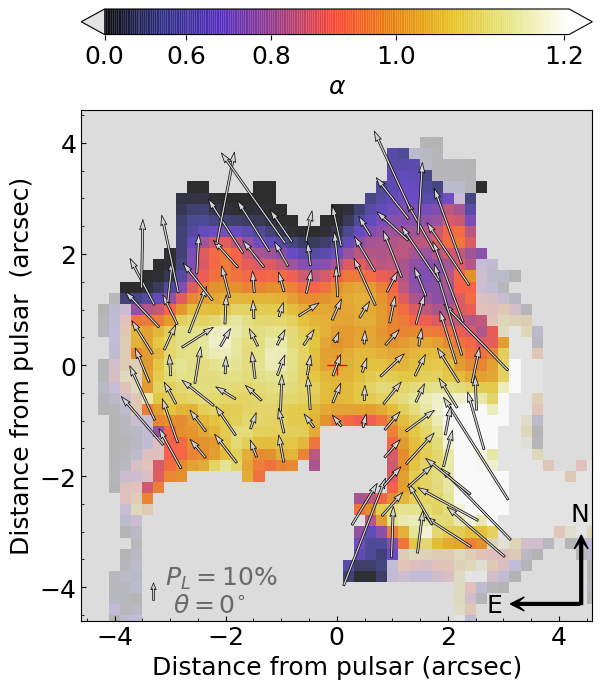}
\caption{Spectral index map of MUSE continuum emission (from \autoref{fig:pixel_map}, bottom right panel, for transparent regions see \autoref{fig:alpha_flux}) compared to \textit{HST} optical linear polarization vectors from \citet{Lundqvist2011}. Arrow length indicates the degree of linear polarization (in $\%$) and the orientation shows the polarization angle.} 
\label{fig:polarisation}
\end{figure}

In \autoref{fig:polarisation}, we overlay the \textit{HST} linear polarization vectors from \citet{Lundqvist2011} onto the MUSE spectral index map from \autoref{fig:pixel_map} (bottom right panel). The polarization is low $\left(\sim 10\%\right)$ in the central parts. We note that along the torus, the direction of polarization is different compared to the rest of the PWN, the angle being approximately perpendicular to the NW-SE axis (although \citealt{Lundqvist2011} report variations of the degree and angle of polarization along the torus on smaller scales). Generally, the degree of polarization increases toward the outer parts of the remnant tracing the spatial hardening of the spectral index.

\subsubsection{\mbox{X-shooter}}
\label{subsubsec:xshooter_whole_slit}

\begin{figure*}[t!]
\plotone{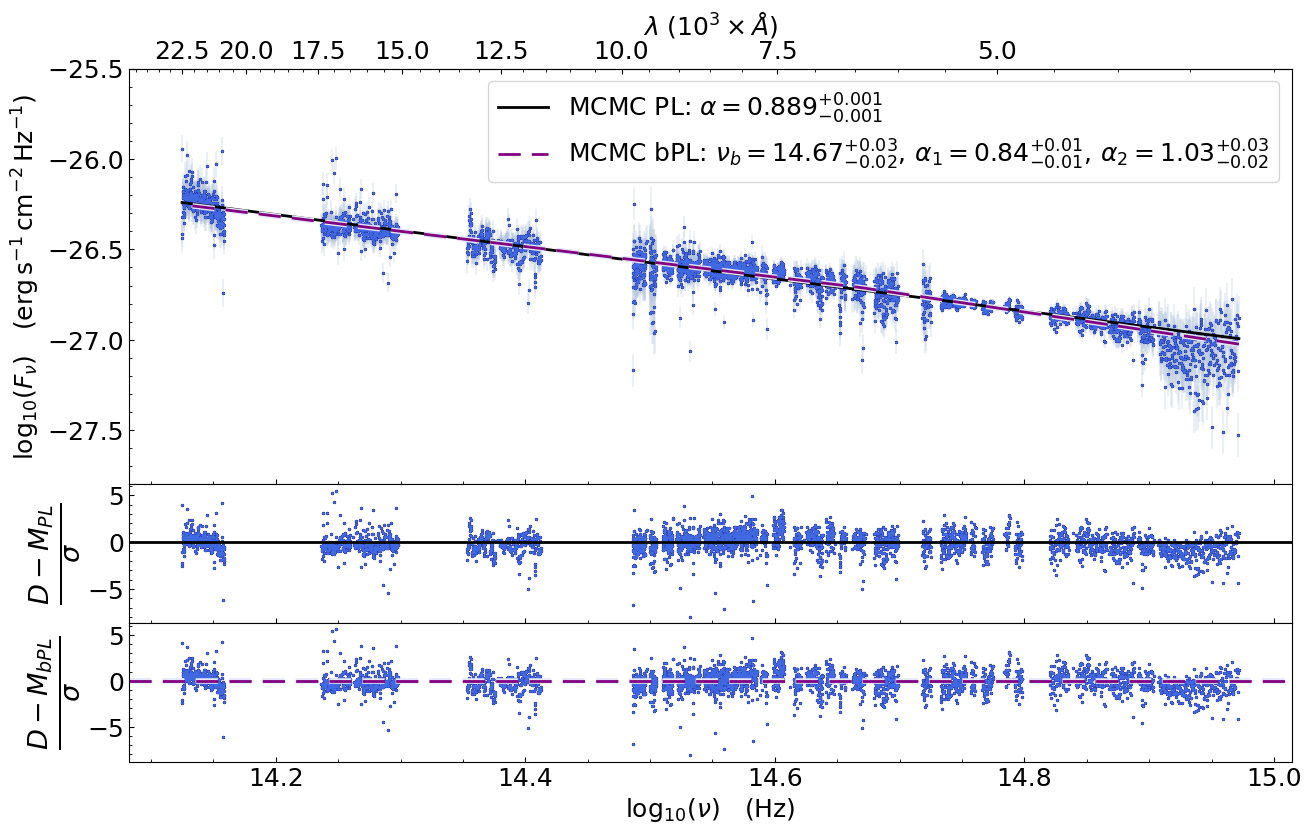}
\caption{\mbox{X-shooter}, PWN: \textit{Top panel:} Continuum emission extracted from the nebula region excluding the pulsar, see \autoref{fig:slit}. The spectrum is fit with a PL model (solid black line) and a bPL model (dashed purple line). \textit{Middle panel:} PL fit residuals computed by subtracting the PL model fit $M_\mathrm{PL}$ from the continuum emission ($D$), which is then dived by the flux uncertainties $(\sigma)$. \textit{Bottom panel:} bPL fit residuals, computed the same way as in the middle panel but using the bPL model $(M_\mathrm{bPL})$.} 
\label{fig:xs_example_fits}
\end{figure*}

Here we study the wider wavelength range from UVB to NIR available with \mbox{X-shooter}. \autoref{fig:xs_example_fits} shows a fit to the \mbox{X-shooter} PWN spectrum extracted from the nebula region, spanning $5\farcs{28}$ around the pulsar, see \autoref{fig:slit}. A single PL fit, that describes the MUSE wavelength range well, is not sufficient when inspecting this wider wavelength range provided by \mbox{X-shooter}. Consequently, we fit the \mbox{X-shooter} spectra with a broken PL (bPL) model 
\begin{equation}
    F_\nu = \begin{cases}
    A\cdot\left(\frac{\nu}{\nu_\mathrm{b}}\right)^{-\alpha_1}, & \text{ if $\nu < \nu_\mathrm{b}$} \\
    A\cdot\left(\frac{\nu}{\nu_\mathrm{b}}\right)^{-\alpha_2},  & \text{ if $\nu > \nu_\mathrm{b}$},
    \end{cases}
\end{equation}
where $A$ is the amplitude, $\nu_\mathrm{b}$ the break frequency and $\alpha_1$ and $\alpha_2$ the low and high frequency spectral indices, respectively. We note that this bPL fit describes the data well, see residuals in \autoref{fig:xs_example_fits} (and fit-results for both PL and bPL models in \autoref{tab:summary_fitresults}), though there are some remaining discrepancies that may be caused by the systematic effects discussed in Section \ref{sec:discussion}. We compare the relative goodness of the single PL and bPL fits with an F-test. The F-test yields a value $>100$ with a p-value of $\ll 0.01$ in favor of the bPL. The best fit bPL model consists of a flatter slope in the lower frequency part of the spectrum (${\alpha_1=0.84^{+0.01}_{-0.01}}$) that breaks at ${\log_{10}\left(\nu_\mathrm{b}\right)=14.67^{+0.03}_{-0.02}}$\,Hz to a slightly steeper slope in the higher frequencies (${\alpha_2=1.03^{+0.03}_{-0.02}}$).

We thus find strong indications for a spectral break in the UVB-NIR wavelength range. There are certain notable caveats related to the NIR part of the spectrum, for example it being less constrained due to the excluded wavelength intervals with poor atmospheric transmission, and having more uncertain flux levels as a result of the different slit widths. More discussion on the \mbox{X-shooter} results and their significance can be found in Section~\ref{subsubsec:PWN:spectral_break}.

\subsection{Spectrum of the pulsar}
\label{subsec:pulsar}

Here we study the pulsar region and hence the emission of the pulsar itself in more detail. By subtracting the nebula contribution (which we call background in this section) from the continuum emission of the pulsar region, we can probe the spectral index of the pulsar itself more accurately. We focus on the MUSE results of the pulsar region. The \mbox{X-shooter} results for the pulsar spectrum are more uncertain and can be found in Appendix~\ref{app_sub:pulsar_systematics}.

\begin{figure*}[t!]
\plotone{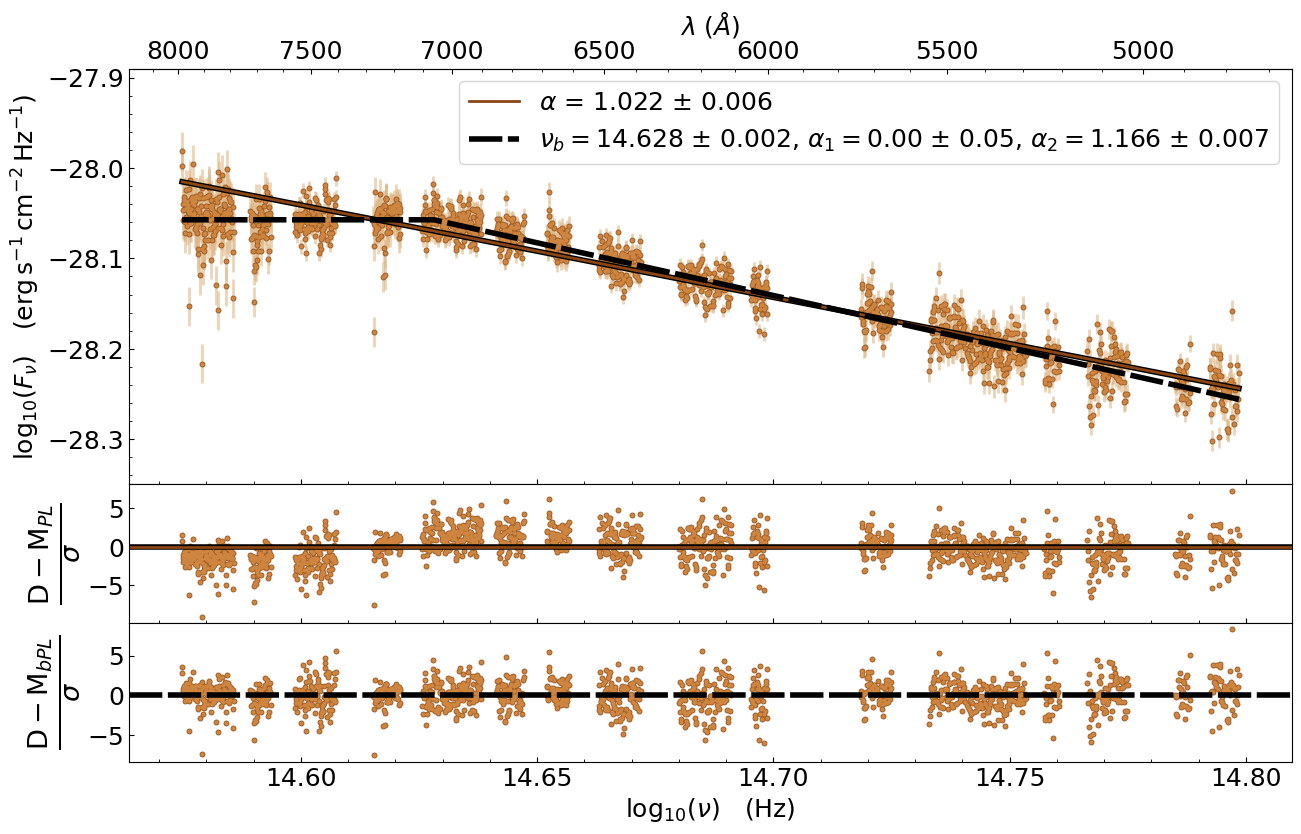}
\caption{MUSE, PSR: \textit{Top panel:} background-subtracted continuum spectrum of the pulsar extracted from a 3$\times$3 spaxel region around the spaxel with the highest flux, see \autoref{fig:slit}. A PL model (solid line) and a bPL model (dashed line) are fit to the continuum. \textit{Middle panel:} PL fit residuals computed by subtracting the PL model fit $M_\mathrm{PL}$ from the continuum emission ($D$), which is then divided by the flux uncertainties $(\sigma)$. \textit{Bottom panel:} bPL fit residuals, computed the same way as in the middle panel but using the bPL model $(M_\mathrm{bPL})$.}
\label{fig:muse_pulsar_continuum_fit}
\end{figure*}

We define the MUSE pulsar region as the region formed by a 3$\times$3 spaxel square (corresponding to 0\farcs{6}\,$\times$\,0\farcs{6}) around the pulsar. The region we use for the background subtraction is an annulus extending between 4 and 6 spaxels (0\farcs{8}, 1\farcs{2}) radii. We subtract the background contribution from the pulsar spectrum and construct the continuum spectrum as described in Section~\ref{sec:construction_of_the_continuum_spectra}. The PWN contribution in the pulsar region is ${\sim 60\%}$, which adds a significant systematic uncertainty to the analysis, especially considering the  spatial variations of the PWN spectrum. The background subtracted pulsar spectrum is shown in \autoref{fig:muse_pulsar_continuum_fit}.

We fit the pulsar spectrum with a single PL model, as shown in the top panel of \autoref{fig:muse_pulsar_continuum_fit}, yielding a spectral index ${\alpha = 1.022 \, \pm \, 0.006}$ (see \autoref{tab:summary_fitresults} for all fit results). Since the single PL fit leaves systematic residuals at low frequencies (where the model overpredicts the data, middle panel of \autoref{fig:muse_pulsar_continuum_fit}), we also fit a bPL to this spectrum. The spectral index at high frequencies, ${\alpha_2= 1.166 \, \pm \, 0.007}$, is similar to the single PL slope, but still significantly different by not overlapping within the formal statistical uncertainty ranges. The lower frequency bPL spectral index is flat ${\alpha_1=0.00 \, \pm \, 0.05}$, and the spectral break is located at ${\log_{10}\left(\nu_\mathrm{b}\right) = 14.628 \, \pm \, 0.002}$\,Hz. These two slopes $\alpha_1$ and $\alpha_2$ are significantly different, and do not overlap even within $5\sigma$ uncertainties. To further quantify the significance of the spectral break, we compare these two models (single PL as the null hypothesis) with the F-test. The F-test statistic is $> 100$ and the corresponding p-value $\ll$\,0.01, showing that the bPL model is a significantly better fit to the pulsar continuum.  

\begin{deluxetable*}{llccccccc}
\tabletypesize{\scriptsize}
\tablewidth{0pt} 
\tablenum{1}
\tablecaption{Summary of the fit results for the PSR and PWN continuum emission. \label{tab:summary_fitresults}}
\tablehead{
\colhead{} & \colhead{} &
\multicolumn2c{PL model} &
\multicolumn5c{bPL model} \\
\colhead{} & \colhead{} & \colhead{$\log_{10}\left(F_\nu\right)^\mathrm{a}$} & \colhead{$\alpha$} & \colhead{$\log_{10}\left(F_\nu\right)^\mathrm{a}$} & \colhead{$\log_{10}\left(\nu_\mathrm{b}\right)$ (Hz)} & \colhead{$\alpha_1$} & \colhead{$\alpha_2$} & \colhead{$\Delta\alpha$}
}
\startdata 
{MUSE} & {} & {} & {} & {} & {} & {} & {} & {}  \\
{4800--8000\,Å} & PSR$^{\rm b}$ & {$-28.02$} & {$1.022\pm0.006$} & {$-28.06$} & {$14.628\pm0.002$} & {$0.00\pm0.05$} & {$1.166\pm0.007$} & {$1.17\pm0.05$}  \\
{} & PWN$^{\mathrm{c}}$ & {$-26.30$} & {$1.033\pm0.004$} & {} & {} & {} & {} & {} \\
{\mbox{X-shooter}} & {} & {} & {} & {} & {} & {} & {} & {} \\
{3200--22,500\,Å} & PWN$^{\rm b}$ & {$-26.57$} & {$0.889^{+0.001}_{-0.001}$} & {$-26.57$} & {$14.67^{+0.03}_{-0.02}$} & {$0.84^{+0.01}_{-0.01}$} & {$1.03^{+0.03}_{-0.02}$} & {$0.19^{+0.03}_{-0.03}$} \\
\enddata
\tablecomments{a) Flux ($\log_{10}\left(F_\nu\right)$, \mbox{erg\,s$^{-1}$\,cm$^{-1}$\,Hz$^{-1}$}) evaluated at $\log_{10}\left(\nu_\mathrm{b}\right) = 14.50$\,Hz. b) Extraction region defined in \autoref{fig:slit}. c) Elliptical extraction region defined in \citet{Mignani2012}.}
\end{deluxetable*}

\section{Discussion} 
\label{sec:discussion}

The shape of the continuum spectra provides information about the particle density and energy distributions as well as the emission mechanisms in SNRs as shown for instance in \citet{Reynolds2017}. In this section we discuss how the first NIR spectroscopy combined with UVB and VIS data contribute to the knowledge of SNR\,0540 and SNRs in general. First, we focus on the PWN\,0540 (Section~\ref{subsec:discussion_PWN}) and discuss its morphology (Section~\ref{subsubsec:PWN_morhpology}), spatial variations of the spectral index (Section~\ref{subsubsec:PWN:spatial_variations}), and the spectral breaks in the continuum spectrum (Section~\ref{subsubsec:PWN:spectral_break}). Second, we turn our attention to the PSR\,0540 (Section~\ref{subsec:discussion_PSR}) and discuss the spectral breaks in the PSR\,0540 continuum emission. In the end, we discuss the results in the framework of PSRs and PWNe in other SNRs in Section~\ref{subsec:comparision_other_SNRs}.

\subsection{PWN} 
\label{subsec:discussion_PWN}

\subsubsection{PWN Morphology} 
\label{subsubsec:PWN_morhpology}

PWNe exhibit axisymmetric structures with equatorial toroidal and polar jet-like features, as is observed e.g for the Crab by \citet{Mori_2004}. Models based on latitude dependence of the pulsar energy flux and magnetization of the pulsar wind have been successful at predicting these kinds of structures (see \citealt{Gaensler2006} and references therein). Previous observations of PWN\,0540 have shown evidence for a torus, detected in optical and X-rays (e.g., \citealt{Morse2003,Gotthelf2000}) and a possible jet (e.g., \citealt{Gotthelf2000,Petre_07}), which our MUSE observations reveal in greater detail.

In the spectral index map (\autoref{fig:pixel_map}, bottom right panel) we clearly see the torus, which is distinctively softer (\mbox{$\alpha\sim1.1$}) than the surrounding emission \mbox{($\alpha\sim$0.6--0.9)}. It spans from NE to SW with an extent of $\sim4\farcs{5}$ and is slightly off of the diagonal NE-SW line. At the LMC distance of $\sim 50$\,kpc, the major axis of the torus is $\sim1.1$\,pc. 

In the convolved flux image (\autoref{fig:pixel_map}, bottom left panel) the torus is not as prominent. A bright region on the SW side can be detected ($\sim80\%$ of the SNR\,0540 maximum flux from the pulsar), but on the NE side the torus is dimmer ($\sim30\%$ of the maximum flux) and blends in with the surroundings. The prominence of the SW part of the torus is even more clear in the deconvolved flux image (\autoref{fig:pixel_map}, top right panel). The SW emission region has a more elongated shape (main emission reaching $\sim$3\farcs{0} away from the pulsar), while, in the NE part, the main emission comes from a smaller region only $\sim$1\farcs{6} from the pulsar. In general in the flux images, the torus seems to follow the diagonal NE-SW direction. 

The torus is most likely composed of magnetic fields and shocked relativistic particles. In X-rays, the torus exhibits a similar or slightly larger size as in the optical, the major axis being $4\farcs{0}-6\farcs{0}$ (e.g., \citealt{Lundqvist2011}), which corresponds to $\sim1.0-1.5$\,pc at the LMC distance. The size of the torus is also comparable in the NIR \citep{Mignani2012} and thus no significant variation in the torus (or general nebula) size is observed in different wavelengths.

Furthermore, the PWN\,0540 torus can be distinguished from the surrounding emission in polarization data as well. The linear polarization angle along the PWN\,0540 torus (pointing NW) differs from the rest of the PWN (pointing NE), as seen in \autoref{fig:polarisation}. Polarization traces the magnetic field configuration and therefore further highlights the difference between the bulk PWN and the torus.

Previous works (\citealt{DeLuca_07} and \citealt{Lundqvist2011}) have observed high brightness variability in the SW parts of the torus. This blob in the SW has been either moving or fading between different epochs (e.g., \citealt{Lundqvist2011}). The deconvolved MUSE fluxes also reveal a bright blob in the SW region ($\sim1$\farcs{8} SW from the pulsar, top right panel of  \autoref{fig:pixel_map}). However, this blob is not clearly separated from the``background'' torus in the convolved MUSE fluxes nor in the \mbox{X-shooter} data. 

The deconvolved MUSE continuum flux image (\autoref{fig:pixel_map}, top right panel) also reveals an arc in the E that is connected to the NE part of the torus. This arc has a radius of $\sim1\farcs{5}$ and spans from the NE to the SE. We find this arc to trace the softest spectral indices in the E (see the flux contours on the bottom right panel of \autoref{fig:pixel_map}) and it also seems to cut through two especially soft regions where $\alpha\sim1.2$. It is possible that this arc could be an artefact originating from the deconvolution procesess, since it is located relatively close to a strong point source (the pulsar), which the deconvolution process enhances. However, we believe that the arc is most likely real, since \citet{Mignani2012} have observed a similar arc in the NIR.

Accompanying the toroidal structure, a jet-like feature in the NW part of central SNR\,0540 has been observed in X-rays by e.g., \citet{Gotthelf2000} and \citet{Petre_07}. Additionally in optical, \citet{Serafimovich_2004} reported possible jets both in the NW and SE. Along with these results, our MUSE observations provide further evidence for a jet. In the MUSE spectral index map (\autoref{fig:pixel_map}, bottom right panel) a region $25^\circ$ from N to the W appears like a jet protruding from the central parts of the torus. This structure features the hardest spectrum in the whole PWN \mbox{($\alpha\sim$0.8--0.5)} although it does not clearly stand out in the flux image. We detect no significant counter-jet in the SE neither in the spectral index nor in the flux image, possibly due to bright field stars occupying this region. Assuming the jet inclination inferred from the 3D morphology is correct ($\sim10^\circ$ away from the observer), the lack of a counter jet cannot be due to relativistic beaming. We also see increased polarization degree in the jet region ($\sim30-40\%$) compared to the central PWN polarization ($\sim10-20\%$), see \autoref{fig:polarisation}.

\subsubsection{Spatial Variations of the Spectral Index within the PWN}
\label{subsubsec:PWN:spatial_variations}

PWN continuum emission can typically be described with PL models over broad wavelength ranges (e.g., \citealt{Mitchell2022}). The slope of the continuum spectrum can vary spatially across the PWN, and the canonical PWN model of fainter outer regions emitting softer spectra stems from the idea of synchrotron cooling. This picture is supported by X-ray observations of the Crab (e.g., \citealt{Mori_2004}), where the spectral index varies from $\alpha\sim0.9$ to $\alpha\sim2$ toward the outer regions. Generally in X-rays, the spectral index seems to soften toward the PWN outer boundary in young SNRs (e.g., \citealt{Hu2022}).

In contrast to this model, we find general spectral hardening in the optical spectra of PWN\,0540, the fainter outer regions being harder (down to $\alpha \sim 0.1$) than the brighter inner parts ($\alpha \sim 1.1$). This spectral hardening toward the PWN outer boundary is further quantified in \autoref{fig:alpha_flux}, where this trend, most prominent in the NE, of dimmer fluxes and lower spectral indices can be seen (diagonal trend below the plateau). However, a local spectral softening toward the outer edge of the PWN in the SW is also observed.

\begin{figure}[t!]
\plotone{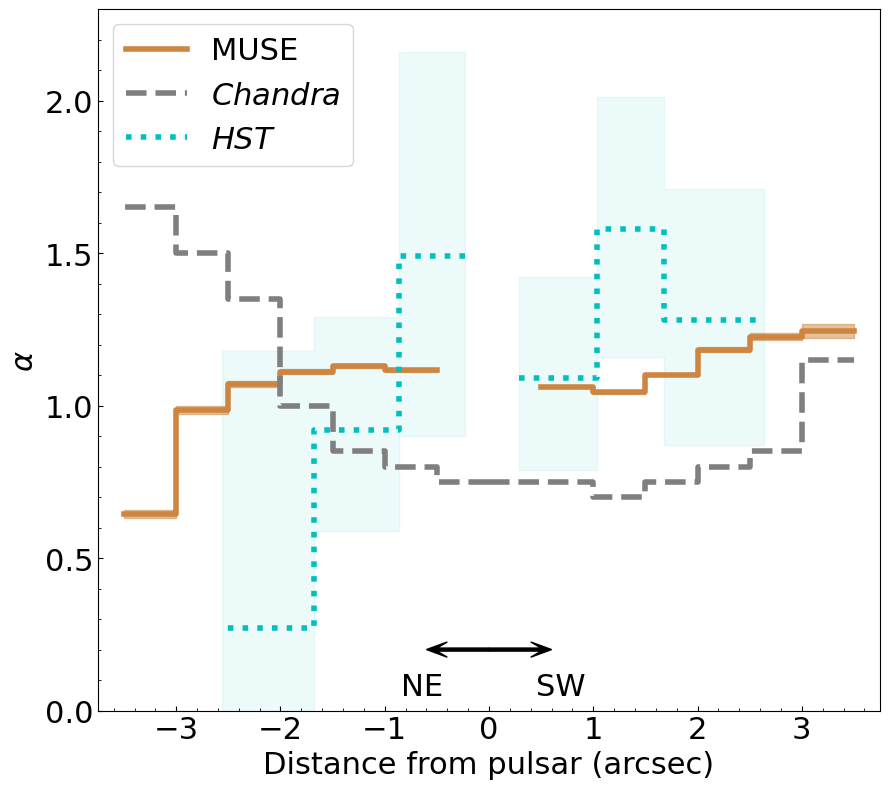}
\caption{Spatial variation of the PWN spectral index $\alpha$ in optical and X-rays. All results are obtained along the torus. MUSE results (this work) are denoted by a golden line. Spectral indices from \citet{Serafimovich_2004} are plotted with a cyan dotted line and \citet{Lundqvist2011} X-ray results with a dashed gray line.}
\label{fig:pwn_alpha_along_torus_comparison}
\end{figure}

Some earlier optical studies of PWN\,0540 have revealed hints of similar spectral hardening toward the outer parts of the PWN. \citet{Serafimovich_2004} used \textit{HST} photometry to measure spectral indices along the NE-SW torus. The findings, also presented in \autoref{fig:pwn_alpha_along_torus_comparison}, show that, in the NE the spectral indices decrease toward the PWN boundary (from $\alpha \sim 1.5$ just NE from the pulsar to $\alpha \sim 0.3$ at $\sim2\farcs{2}$ NE from the pulsar). We compare our MUSE results to these values by extracting continuum spectra similarly along the torus. We find that the spectral indices from \citet{Serafimovich_2004} mostly overlap with our MUSE results, as shown in \autoref{fig:pwn_alpha_along_torus_comparison}. However, the measurement uncertainties were too large for \citet{Serafimovich_2004} to confirm the possible spatial variation. No other previous measurement in the optical provides as detailed information on spatial variations of the spectrum. \citet{Mignani2012} report minor evidence for spectral hardening away from the pulsar in the NIR and optical by obtaining a softer spectral index (${\alpha=0.70\pm0.04}$) for PSR\,0540 and a harder spectral index (${\alpha=0.56\pm0.03}$) for the whole spatially averaged PWN\,0540.

In X-rays, however, observations show radial softening of the spectral index toward the PWN\,0540 outer regions. \citet{Petre_07} used \mbox{\textit{Chandra}} X-ray observations to map the radial variation of the spectral index by using concentric elliptical annuli as extraction regions. They find that the pulsar (which is strongly affected by pile-up) completely contaminates the innermost region within $\sim1\farcs{0}$ radius, but that there is a softening of spectral indices from $1\farcs{0}$ toward the outer regions of the nebula (from ${\alpha\sim0.4}$ to ${\alpha\sim1.4}$). Beyond $\sim5\farcs{0}$, they conclude that a more complex model than a single PL is needed to explain the emission in X-rays.

Also \citet{Lundqvist2011} derived an X-ray spectral index map for PWN\,0540 from \textit{Chandra} data with some emphasis along the torus. Similar to the MUSE and \mbox{X-shooter} results along the torus in \autoref{fig:pixel_map} and \autoref{fig:xs_example_fits}, where $\alpha\sim1.1$ and $\alpha_2\sim1$, respectively, \citet{Lundqvist2011} find an approximately constant spectral index in X-rays ($\alpha\sim0.8$) within the $\sim2\farcs{0}$ radius both to the SW and NE of the pulsar, see \autoref{fig:pwn_alpha_along_torus_comparison}. Beyond $\sim2\farcs{0}$ and up to $\sim5\farcs{0}$ distance from the pulsar, they observe softening of the spectral index ($\alpha\sim 0.8-1.8$) both to the NE and SW. Thus, comparing the optical and X-ray results reveals that the significant variation (softening in X-rays and hardening in optical) begins at $\sim2''$ radius. The only exception to this picture is the SW part of the torus, where the optical spectral index softens further out (\autoref{fig:pixel_map} and \autoref{fig:pwn_alpha_along_torus_comparison}). This particular section of the PWN  therefore seems to obey the simple synchrotron cooling picture. However, the flux from this region is low and it is located near the excluded high-$\alpha$ region, which might hint at systematic effects affecting the spectral index results.

The origin of the apparent divergence from the simple synchrotron cooling picture for most of the nebula in the optical is not clear. One consideration is that the convolution (Section \ref{sec:construction_of_the_continuum_spectra}) does not perfectly correct for the MUSE PSF wavelength dependence. While this introduces a systematic uncertainty in the spectral index map, it is not expected to affect the main conclusions, considering that the spectral index hardening is so prominent (the difference between the inner and outer spectral indices being ${\Delta\alpha\sim1}$). 

Another reason behind this deviation from the simple synchrotron cooling picture could be that there is an additional continuum emission component other than synchrotron that contributes significantly to the optical spectrum. The most likely such contribution is Balmer recombination continuum and 2-photon emission, which may be expected at a low level, similar to the case in the Crab nebula \citep{Veron-Cetty1993}.  

However, in this case we expect a spatial correlation between the hard (blue) continuum and the Balmer emission lines in the central part of the remnant, which is not observed. In particular, the MUSE data reveal weak H$\beta$ emission with a similar distribution as the [\ion{S}{3}] emission in \autoref{fig:linemap_vs_continuum}, but with the most pronounced emission in the northern part of the SW ring structure (see Figure 1 in L21). By contrast, the hardening of the optical continuum emission is observed in the outer regions all around the PWN except the SW. Interestingly, the polarization properties shows a similar systematic variation (\autoref{fig:polarisation}). 

Assuming that the optical continuum is indeed dominated by synchrotron emission, the hardening towards the outer regions may be due to reacceleration of particles by the pulsar wind shock at the inner edge of the ejecta, though we note that such a scenario lack theoretical predictions. This seems to be consistent with the relative location of the continuum and line emitting regions (Figure\,\ref{fig:linemap_vs_continuum}), though the interpretation is clearly complicated by projection effects. Another possibility may be that the hardening is a result of time variability of the pulsar wind. In this scenario, the wind further from the pulsar would have been produced at an earlier time, possibly with different properties, which would then result with spatially variable spectral index.

The main caveat affecting the comparison between the optical and X-rays above is the possible time variability, observed e.g., by \citet{Petre_07}, which makes comparing results from previous observations challenging. In addition, the results may also be affected by line-of-sight effects, where emission, say from the central PWN, is mixed with PWN contributions further away that reside in the fore- or background along the line-of-sight. This is complicated to account for since PWN\,0540 is known to have a highly asymmetric three dimensional morphology (e.g., \citealt{Sandin2013} and L21).

\subsubsection{The PWN Multiwavelength Spectrum} 
\label{subsubsec:PWN:spectral_break}

After considering the high spatial resolution MUSE observations, we turn our focus to the wider spectral range provided by \mbox{X-shooter}. We find strong indications that a single PL model is not sufficient to characterize the PWN\,0540 continuum emission in the UVB-NIR ($10^{14}$--$10^{15}$ Hz) range, and instead fit the continuum emission with a bPL model. In this section, we discuss the implications of these \mbox{X-shooter} results and how they fit into the multiwavelength context from the radio to X-rays.

The main challenge with the X-shooter observations is the connection of the three spectra: NIR, VIS and UVB. It is clear that the NIR part of the X-shooter spectrum is overall less constrained because of the large gaps in the data (due to atmospheric transmission that affects all ground-based IR observations) and therefore it is especially challenging to connect the NIR and VIS bands perfectly. These challenges add a systematic uncertainty to all spectral index results, which we estimate to be at least $0.1$ (for a more detailed discussion, see \ref{app_sub:agreement}). Additional challenges come with the limited field of view of the slit i.e. it does not cover the entire PWN.

We can study the details of the synchrotron emitting particles by focusing on the spectral index difference between the low and high frequencies, $\Delta\alpha \equiv \alpha_2 - \alpha_1$. According to the simplest synchrotron model with constant energy injection and with a spatially uniform PL distribution for the emitting particles, the difference between the injected (higher frequency) and cooled (lower frequency) spectral slopes is $\Delta\alpha = 1/2$. The difference between the best-fit low and high-frequency spectral index within the \mbox{X-shooter} spectrum in \autoref{fig:xs_example_fits} is $\Delta\alpha=0.19^{+0.03}_{-0.03}$, which significantly deviates from the canonical value 1/2. Generally, since PWN\,0540 is clearly not spatially uniform (as is dicussed in Sections \ref{subsubsec:PWN_morhpology} and \ref{subsubsec:PWN:spatial_variations}), the exact value of 1/2 is not expected. Additional factors coming from the aforementioned systematic effects at play in the X-shooter fits might also affect the difference of the two spectral indices. However, this $\Delta\alpha$ result deviates from the canoncial value by more than $5\sigma$, which indicates that the break between the NIR and UVB is most likely not caused by synchrotron losses.

Before discussing how the MUSE and \mbox{X-shooter} UVB-NIR results connect to other wave bands, we address caveats that arise from time and spatial variability of the PWN. During 2011, PSR\,0540 experienced a sudden increase in its spin-down rate \citep{Marshall_15}, which also caused an increase in the PWN X-ray flux by $\sim$\,32\%, reported by \citet{Ge_19}. It is believed that a change in the pulsar magnetosphere caused the PWN brightening and the X-ray flux values have remained at the same elevated level after the gradual increase between late 2011 and 2016 (the latest observation performed in 2019). In addition, \citet{DeLuca_07} and \citet{Lundqvist2011} have observed changes in the spatial distribution of the optical continuum emission, especially in the SW, during relatively short time scales of $\sim$\,10 yr. These observed changes in X-ray flux levels and optical (spatial) continuum distribution introduces uncertainties in the comparison between the recent observations to the pre spin-down rate change observations or observations performed $\gtrsim$ 10 years ago.

\begin{figure*}[t!]
\plotone{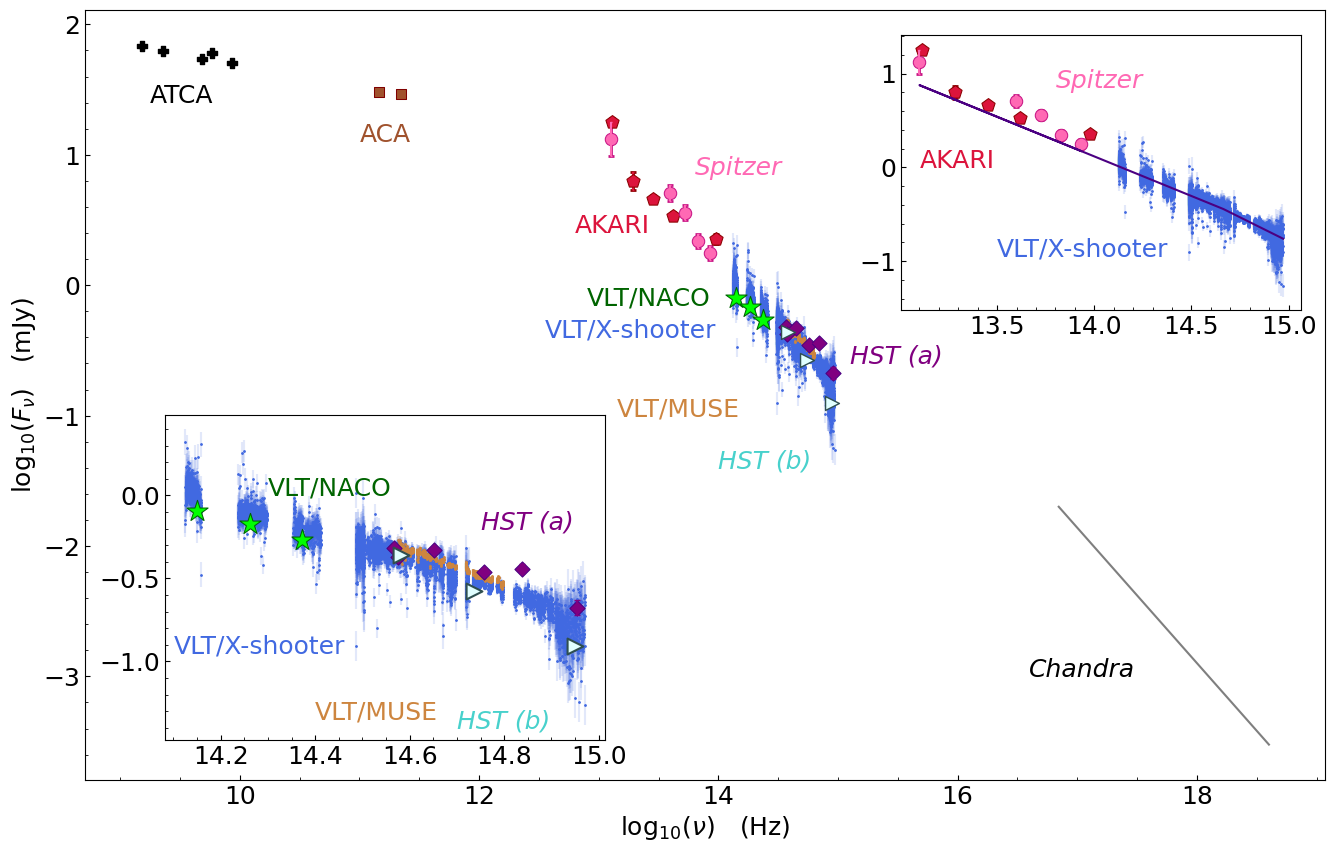}
\caption{Spatially averaged multiwavelength continuum spectrum of PWN\,0540. The VLT/MUSE (golden points) and VLT/\mbox{X-shooter} (blue points) are from this work. The VLT/MUSE data are acquired through the same extraction region as is used in \citet{Mignani2012}. The VLT/\mbox{X-shooter} extraction region is shown in \autoref{fig:slit} as the nebula region. For the X-shooter flux-level calibration, see text. Only MUSE and X-shooter data are corrected for extinction with the new color excess value ${E(B-V)=0.27}$. In the radio, the ATCA data (black plus-signs) are from \citet{Brantseg2014} and ACA data (brown squares) from \citet{Lundqvist2020}. The IR regime is covered by the \textit{Spitzer} data (pink circles) from \citet{Williams_2008} and the AKARI data (red pentagons) from \citet{Lundqvist2020}. The NIR range is covered by data from \citet{Mignani2012} (green stars) and the two optical \textit{HST} observations are from \citeauthor{Mignani2012} (\citeyear{Mignani2012}, \textit{HST} a, purple diamonds) and \citeauthor{Serafimovich_2004} (\citeyear{Serafimovich_2004}, \textit{HST} b, light blue triangles). The X-ray data are represented with a slope of $\alpha_{X}\sim1$ (gray solid line) and are from \citet{Kaaret_2001}. If uncertainties are not presented, they are smaller than the used markers. \textit{Bottom Inset:} A zoom-in of the UVB-NIR range. \textit{Top Inset:} A zoom-in to the UV-IR range. A bPL fit (purple solid line) to the UVB-NIR \mbox{X-shooter} data is presented with the IR data.}
\label{fig:pwn_multi}
\end{figure*}

In \autoref{fig:pwn_multi} we present spatially averaged results of the PWN\,0540 multiwavelength continuum spectrum. The plotted MUSE spectrum is extracted through the same elliptic extraction region ($3\farcs{4}\times 2\farcs{4}$ centered at the pulsar but the pulsar excluded) that is used in \citet{Mignani2012}. The resulting PL fit gives ${\alpha=1.033\pm0.004}$. For \mbox{X-shooter}, we used the nebula region defined in \autoref{fig:slit}. We then scaled the flux levels to correspond to the flux levels of the larger \citet{Mignani2012} extraction region, using MUSE to determine the flux ratio between the two regions. As can be seen in \autoref{fig:pwn_multi}, the \mbox{X-shooter} and MUSE spectra are overlapping and give similar results for the spectral indices within $1\sigma$ (see also \autoref{tab:summary_fitresults}). We note, however, that this nearly perfect agreement might be serendipitous since the extraction regions are different and there are known systematic differences between the instruments (Appendix~\ref{app:muse_vs_xshooter})

Previous results in the optical and NIR by \citet{Serafimovich_2004} and \citet{Mignani2012} mostly overlap with the \mbox{X-shooter} and MUSE results. Differences in the slopes and flux levels may occur due to different extraction regions and observational epochs. Additionally, many of the photometric results (e.g., \citealt{Mignani2012}) seem to be contaminated by emission lines, for instance with the [Fe II] 1.26 $\mu$m line in the NIR and especially [\ion{O}{3}]\,$\lambda\lambda$4959,\,5007 in the optical. In conclusion, all these challenges explain the conflicting results previously obtained for the spectral index in the optical range (\citealt{Serafimovich_2004,Mignani2012}).

As can be seen in \autoref{fig:pwn_multi}, previous works have reported higher flux densities in the IR, especially in the lower frequency part (\textit{Spitzer} and AKARI results, \citealt{Williams_2008} and \citealt{Lundqvist2020}) compared to the optical (VLT/NACO and \textit{HST} results, \citealt{Mignani2012} and \citealt{Serafimovich_2004}). A dust component with $T\sim50$\,K and $M_{\mathrm{dust}}$\,$\sim$\,$3\cdot10^{-1}$\,$\mathrm{M}_\odot$ has been proposed to account for this excess emission at \mbox{log-frequency}\,${\sim13}$\,Hz (\citealt{Williams_2008} and \citealt{Lundqvist2020}). \citet{Lundqvist2020} computed a spectral slope of $\alpha_\mathrm{IR} = 0.87^{+0.10}_{-0.08}$ between the far IR dust bump and the NIR/optical.

The \mbox{X-shooter} spatially averaged nebula continuum seems to fit this picture. In the top right inset of \autoref{fig:pwn_multi}, we show how the lower frequency spectral index ${\alpha = 0.84^{+0.01}_{-0.01}}$ obtained from a bPL fit to the \mbox{X-shooter} data connects the NIR and IR emission. However, this good agreement may be a coincidence due to the time variability discussed above, in combination with likely contamination by lines in the IR (discussed in \citealt{Lundqvist2020}), and the different color excess values used. We therefore conclude that dust emission in the IR is possible but cannot be confirmed with our data.

If we assume that the IR fluxes at \mbox{log-frequency}\,${\sim13}$\,Hz are contaminated by line emission, we can study how the UVB-NIR fluxes are connected to the radio band. Since the optical-NIR spectral index ${\alpha = 0.84^{+0.01}_{-0.01}}$, see fit in \autoref{fig:pwn_multi}) aligns well with the corresponding value from \citet{Lundqvist2020} ($\alpha=0.87^{+0.10}_{-0.08}$), we end up with the same conclusion that there has to be at least one spectral break between the IR and the radio, in addition to the spectral break we found between the NIR and UV (at \mbox{$\log_{10}\left(\nu_\mathrm{b}\right)=14.67^{+0.03}_{-0.02}$\,Hz}). 

For the spectral break between radio and IR wavelengths, \citet{Lundqvist2020} obtains a location at the \mbox{log-frequency}\,$\sim12.7$\,Hz with the radio spectral index being $\alpha = 0.17\pm0.02$. Similar results were reached by \citet{Brantseg2014}, where the spectral break is located at the \mbox{log-frequency}\,$\sim13.8$\,Hz with a radio spectral index of $\alpha = 0.16\pm0.1$. The magnitude of the break is consistent with a cooling break.

On the higher frequency side, \citet{Kaaret_2001}\footnote{The spectrum from \citet{Kaaret_2001} is currently the only X-ray spectrum of the full integrated PWN, which also lists the fluxes, that is available in the literature.} measured the PWN continuum spectrum in X-rays and found the spectral index $\alpha=1.04\pm0.18$. This result lies within the uncertainty range of the \mbox{X-shooter} spatially averaged higher frequency spectral index, ${\alpha_2=1.03^{+0.03}_{-0.02}}$, implying that no break between the optical and X-rays would be required. However, the picture is most likely more complicated.  In addition to the larger color excess value used in this work, the X-ray fluxes measured by \citet{Kaaret_2001} overshoot the optical fluxes, as can be seen in \autoref{fig:pwn_multi}. In reality, the situation is even worse, the overshooting being enhanced by the $32\%$ increase in X-ray fluxes (\citealt{Ge_19}).

In conclusion, the evidence for multiple breaks (a cooling break between the log-frequencies \mbox{12--13}\,Hz and another break between \mbox{14--15}\,Hz) supports the scenario of multiple particle populations at play in the PWN\,0540, unless the multiwavelength comparison is strongly affected by time variability.

\subsection{Pulsar} 
\label{subsec:discussion_PSR}

\subsubsection{PSR Multiwavelength Spectrum} 
\label{subsubsec:PSR_spectral_break}

MUSE observations show that the pulsar spectrum in the optical exhibits a similar spectral index (${\alpha\sim1}$) to the central PWN spectrum (bottom right panel of \autoref{fig:pixel_map}). However, the statistically preferred model for the pulsar was found to be the bPL model. The bPL best-fit model has a break frequency at $\log_{10}\left(\nu_\mathrm{b}\right)$\,$=$\,$14.628$\,$\pm0.002$\,Hz, with spectral indices ${\alpha_1 = 0.00 \pm 0.05}$ for the lower frequency and ${\alpha_2 = 1.166 \pm 0.007}$ for the higher frequency part of the spectrum. 

Previous observations in the optical have yielded slightly harder (smaller $\alpha$) optical spectral indices for PSR\,0540, and no evidence for a spectral break has been reported in this wavelength range. Spectroscopic observations of PSR\,0540 in the optical, performed by \citet{Serafimovich_2004}, resulted in the spectral index ${\alpha = 1.07^{+0.20}_{-0.19}}$ (or ${\alpha = \left[1.07-0.2\right]^{+0.20}_{-0.19}}$ to counter the larger color excess value used in this work). \citet{Mignani2012} used photometry in the \mbox{optical-NIR} and report an even harder spectral index of ${\alpha = 0.70 \pm 0.04}$. A general hardening in the NIR range is compatible with our results, though the values of $\alpha$ clearly differ. 

As discussed in Section~\ref{subsec:discussion_PWN}, there are many factors that make it challenging to compare our results to the ones provided in the literature. The most significant of these is the spin-down rate change in 2011 \citep{Marshall_15} related to which the PWN X-ray flux levels were measured to have increased by $\sim30\%$, but the PSR flux in X-rays appear to have stayed unchanged \citep{Ge_19}. How the spin-down rate change has affected pulsar fluxes in lower energies, for example in the optical range, has not been studied. 

To our knowledge, \citet{Mignani2019} have performed the only observations after the 2011 spin-down rate change that focus on the PSR\,0540 spectrum. They find that the near and far UV (NUV and FUV, bands around 2350 and 1590\,\AA, respectively) fluxes are completely incompatible with the pre-2011 optical spectra. The spectral index they infer is ${\alpha\sim3}$, the largest spectral index ever recorded for a pulsar in the UV. Thus, \citet{Mignani2019} deduce that a turnover below $\sim2400$\,\AA\,(or in log-frequencies \mbox{above\,$\sim15$\,Hz}) is needed to explain the optical-UV spectrum. Our optical spectral index $\alpha\sim1$ follows the spectral index trend reported for other pulsars in the optical \mbox{$\alpha\sim$\,0--1} \citep{Mignani2007}, but is incompatible with the UV slope from \citet{Mignani2019}. We are not able to draw any conclusions regarding a possible spectral turn-over around the \mbox{log-frequency\,15 Hz} due to a high noise level in the \mbox{X-shooter} spectra at these frequencies. 

On the other hand, \citet{Mignani2019} assumed the color excess value of ${E(B-V)=0.20}$\,mag, which complicates the comparison with our results. Taking the color excess value found in this work, ${E(B-V)=0.27}$\,mag, and applying it to the results reported in \citet{Mignani2019}, we find that the UV spectral index remains at the high value $\alpha\sim3$. This is due to the shape of the extinction curve (2175\,Å UV bump), which does not change significantly when varying the color excess values.

In X-rays, the PSR\,0540 spectrum seems to be harder than in the optical, at least according to pre-2011 observations. \citet{Lundqvist2011} report ${\alpha=0.74\pm0.01}$ for the PSR\,0540 X-ray spectral index whereas \citet{Kaaret_2001} report ${\alpha=0.92 \pm 0.11}$. The latter is still $\sim3\sigma$ from our MUSE higher frequency spectral index ${\alpha_2 = 1.166 \pm 0.007}$. We confirm the previous conclusion \citep{Serafimovich_2004,Mignani_2010} that, even without the \citet{Mignani2019} results, the PSR\,0540 spectral energy distribution from the optical to the X-rays cannot be described with a single PL.

\subsection{Comparison to the Crab Nebula and pulsar}
 \label{subsec:comparision_other_SNRs}

In this section, we compare the main results of PSR\,and\,PWN\,0540 from this work to the most similar SNR, the Crab Nebula. Our MUSE results confirm that the morphology and size of the PWN\,0540 torus resemble the torus of the Crab (see earlier comparison in X-rays by \citealt{Gotthelf2000}). For the Crab torus the major axis of the outer ring is $\sim\,2\,\times\,41''$ in X-rays (projected onto the plane of the sky, \citealt{Ng2004}, see also \citealt{Mori_2004}) which at the $\sim 2$ kpc distance \citep{Hester2008} translates to $\sim 1$\,pc. This is interestingly the same length as the torus in SNR\,0540. The size of torus in PWN\,0540 stays approximately constant from NIR to X-rays, in contrast to the Crab Nebula, which increases in size at longer wavelengths (see e.g., \citealt{Hester2008}).

As mentioned before, work by \citet{Veron-Cetty1993} on the Crab Nebula is one of the few studies on the PWN spectral index spatial variations in optical wavelengths. Their results indicate spatial spectral softening toward the PWN outer edges, whereas our MUSE observations show the opposite for PWN\,0540. 
The Crab results are, however, not directly comparable because they are based on narrow band photometry at a few wavelengths, unlike the MUSE results that build on an entire spectrum.

In addition to the differences in the PWN emission, we can also compare the PSRs in SNR\,0540 and Crab Nebula. \citet{Sollerman2019} studied emission from the Crab pulsar with \mbox{X-shooter} and found no strong indications for a spectral break in the UVB-NIR range. More interestingly, the Crab pulsar’s spectral index seems to have the opposite sign, $\alpha=-0.16\pm0.07$ compared to PSR\,0540 (as already noted in \citealt{Mignani2007}), despite their similar ages and energetics. This might indicate that the differences in the pulsars at least partly propagate to the PWNe causing the opposite trend in the spatial spectral index variation. 

The Crab pulsar's birth spin period is \mbox{$\sim17$\,ms} \citep{Lyne2015}. In comparison, we obtain the birth spin period ${P_0=31.70\pm0.04}$\,ms for PSR 0540 by applying the new age estimate given by L21 and the most recent braking index measurement by \citet{Ge_19}. A more detailed discussion can be found in Appendix \ref{app:pulsar_age}.

\section{Summary and Conclusions} 
\label{sec:conclusions}

We have studied the continuum emission from PSR\,and\,PWN\,0540 in the SNR\,0540 using the VLT instruments MUSE and \mbox{X-shooter}. The spectra cover the wavelength range \mbox{4650–9300\,\AA} in the optical (MUSE) and \mbox{3000--25,000\,\AA} in the UVB-NIR (X-Shooter). We mapped the Balmer decrement in regions close to SNR\,0540 and determined the color excess ${E(B-V)=0.27}$\,mag. This value is larger than previous findings (\citealt{Kirshner_1989,Serafimovich_2004}) and affect the spectral index results by flattening the spectra (i.e. decreasing $\alpha$) by ${\sim0.2}$. We fit the continuum spectra with PL and bPL models in order to study the shape of the spectra. The main conclusions are summarised below.

\begin{enumerate}
    \item We can identify several structures within the PWN by forming a spectral index map of the best-fit PL spectral indices with MUSE. First, we confirm the torus-like structure, exhibiting approximately constant spectral index of $\alpha\sim1.1$, spanning from the NE to the SW with a major axis\,$\sim4$\farcs{5}. The continuum flux map shows a similar elongation, with the SW part being significantly brighter. Second, we see evidence for a possible jet in the NW, where the spectra are distinctively harder \mbox{($\alpha\sim$0.5--0.9)}. No such structure is seen in the flux map. Additionally, we do not find evidence for a counter-jet in the SE. Third, after decovolving the MUSE image, we identify an arc, which is connecting to the torus in the NE and spanning about half a circle to the S with a radius of\,\mbox{$\sim1$\farcs{5}}.
    
    \item We find clear evidence of spectral hardening of the optical spectrum toward the outer edges around most of the PWN (from $\alpha\sim1.1$ to $\alpha\sim0.1$). This is opposite to the trend observed in X-rays and to the expectation from synchrotron losses. Comparison with previous polarization data also shows variations on similar spatial scales. The origin of the spectral hardening is not clear, but may indicate re-acceleration of particles at the inner edge of the ejecta or time-variability of the pulsar wind.

    \item The PL fit to the spatially integrated spectrum of the PWN in MUSE continuum spectrum give a spectral index $\alpha=1.033\pm0.004$.  When fitting the full UVB-NIR spectrum, available along the torus covered by the \mbox{X-shooter} slit, we find that the PWN\,0540 continuum spectrum can be better characterised with a bPL. The spatially averaged bPL best fit results locate the spectral break at ${\log_{10}\left(\nu_\mathrm{b}\right)=14.67^{+0.03}_{-0.02}\,\mathrm{Hz}}$ with the spectral indices ${\alpha_1=0.84^{+0.01}_{-0.01}}$ and ${\alpha_2=1.03^{+0.03}_{-0.02}}$ at lower and higher frequencies, respectively. A comparison of these results with previous multiwavelength observations shows that several breaks are needed to explain the full radio-X-ray spectral energy distribution, which likely indicates more than one population of particles. 
    
    \item The PSR\,0540 spectrum observed by MUSE exhibits a spectral break at the \mbox{log-frequency}\,${\log_{10}\left(\nu_\mathrm{b}\right)=14.628\pm0.002}$\,Hz and can be described with the spectral indices ${\alpha_1=0.00\pm0.05}$ (lower frequencies) and ${\alpha_2=1.166\pm0.007}$ (higher frequencies). This implies that the pulsar spectrum has a similar spectral index ($\alpha_2$) to the torus of the PWN. The spatial resolution of MUSE allows us to isolate the pulsar spectrum better than in previous works, though we note that the result is still affected by systematic uncertainties due to the high and spatially varying background from the PWN.
\end{enumerate}

Due to the time variability of SNR\,0540 and its pulsar and PWN, it is challenging to interpret the full spectral energy distribution when there are no recent simultaneous multiwavelength observations. For this reason, quasi-simultaneous observations, for example with \textit{JWST} in the IR and \textit{Chandra} in the X-rays, would be needed to draw stronger conclusions on the shape and break locations of the PSR and PWN\,0540 continuum spectra. 
Further work is also needed to understand the origin of the spatial spectral index variations. There are very few such studies in the optical, and, to the best of our knowledge, PWN0540 is the only case where spectral hardening toward the edges is seen. Similar studies of other PWNe are and theoretical modeling would therefore be of great interest. 


\begin{acknowledgments}
JL acknowledges support from the Knut \& Alice Wallenberg Foundation. PL would like to thank the Swedish Research Council for support. JDL acknowledges support from a UK Research and Innovation Fellowship (MR/T020784/1). We thank Sophie Rosu for providing the HST image.
\end{acknowledgments}

%

\vspace{5mm}
\facilities{VLT (MUSE and \mbox{X-shooter}).}

\software{Astropy \citep{astropy13,astropy18,astropy22}, emcee \citep{emcee2013}, matplotlib \citep{matplotlib}, photutils \citep{photutils}, scikit-image \citep{scikit-image}, scipy \citep{scipy}.}



\appendix

\section{Comparison of MUSE and \mbox{X-shooter} spectra}
\label{app:muse_vs_xshooter}

\subsection{Comparison of the Two Instruments}
\label{app_sub:agreement}

The fact that MUSE and \mbox{X-shooter} have different spatial and spectral resolution, different pixel sizes, seeing, and calibration uncertainties can cause systematic effects when analysing the spectra. In addition, the placement and dimensions of the \mbox{X-shooter} slit and the different slit widths naturally limit the extraction region. In this section we cross-check the agreement between the two instruments.

\begin{figure*}[t!]
\plotone{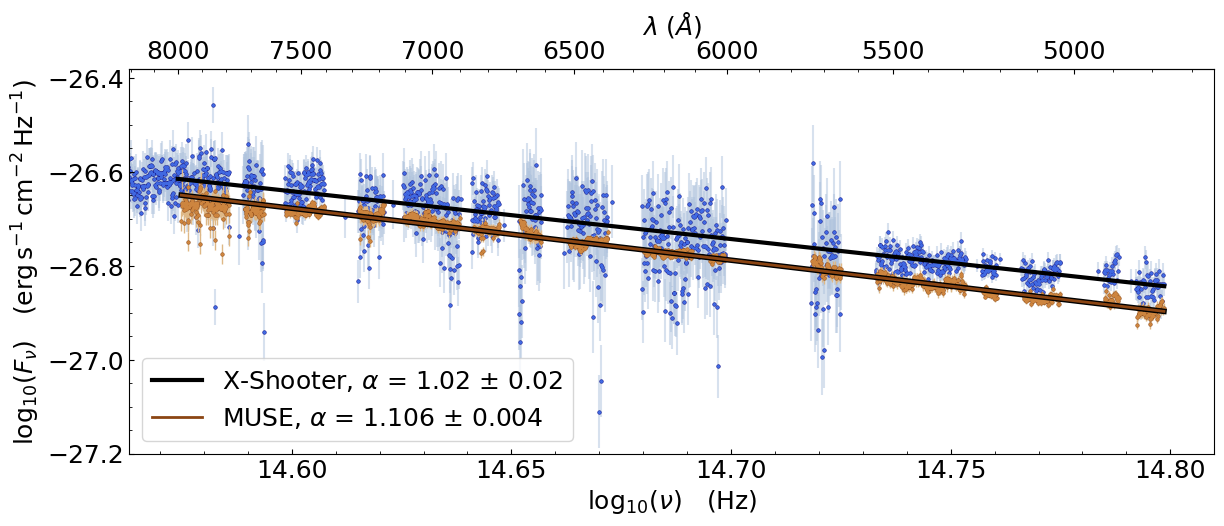}
\caption{Comparison of PWN spectra from \mbox{X-shooter} (blue points) and MUSE (brown points). The MUSE spectrum is extracted with a pseudo-slit resembling the \mbox{X-shooter} slit (constrained by the cyan lines in \autoref{fig:slit}). The two lines (black for \mbox{X-shooter} and brown for MUSE) represent PL fits to these spectra. Note that the \mbox{X-shooter} spectrum is binned.}
\label{fig:appendix_comparison_nebula_muse_vs_xs}
\end{figure*}

We start by studying the PWN spectrum with \mbox{X-shooter} and MUSE, both extracted through the same \mbox{X-shooter} slit (or a 1\farcs{6} wide pseudo-slit for MUSE, see cyan lines in \autoref{fig:slit} indicating the nebula region). We correct the \mbox{X-shooter} spectra for the different slit widths by scaling the NIR and VIS fluxes according to the slit width ratios using the UVB slit as a reference. We fit (with the least-squares method) a PL model for both spectra in the MUSE range, shown in \autoref{fig:appendix_comparison_nebula_muse_vs_xs}, with the resulting spectral indices $\alpha=1.02\pm0.02$ and $\alpha=1.106\pm0.004$, for \mbox{X-shooter} and MUSE, respectively. Since these two spectra are extracted through the largest possible common extraction region and fitted over the full overlapping wavelength range, it provides us a way to measure the underlying systematics. The difference between the \mbox{X-shooter} and MUSE spectral index is $\sim0.1$, which is significantly larger than the formal statistical fit uncertainties. We thus set the systematic offset to be at least $0.1$ in all measurements.

As the pulsar region is of particular interest, we continue studying the systematic effects by doing the same exercise as above but, now for the pulsar region by extracting spectra through the \mbox{X-shooter} slit (a pseudo-slit for MUSE) around the pulsar. We define the pulsar region as the area enclosed by the black lines in \autoref{fig:slit}. This region is 0\farcs{96} wide and extends $0\farcs{4}$ to the NE and $0\farcs{56}$ to the SW from the pulsar center. The width 0\farcs{96} was chosen so that the majority of the pulsar emission would be included. However, this definition for the pulsar region is almost three times bigger than the corresponding pulsar region defined for the MUSE analysis (see Section~ \ref{subsec:pulsar}), which is necessary due to the different spatial resolutions and sampling between the two instruments. Additionally, we avoid introducing additional factors to this pulsar region analysis by not applying background subtraction to the spectra. Because of this, the emission from this region is dominated by the PWN, and we correct the \mbox{X-shooter} spectra for the slit widths in the same way as above. The effect of the background subtraction is studied in Appendix~\ref{app_sub:pulsar_systematics} below.

\begin{figure*}[t!]
\plotone{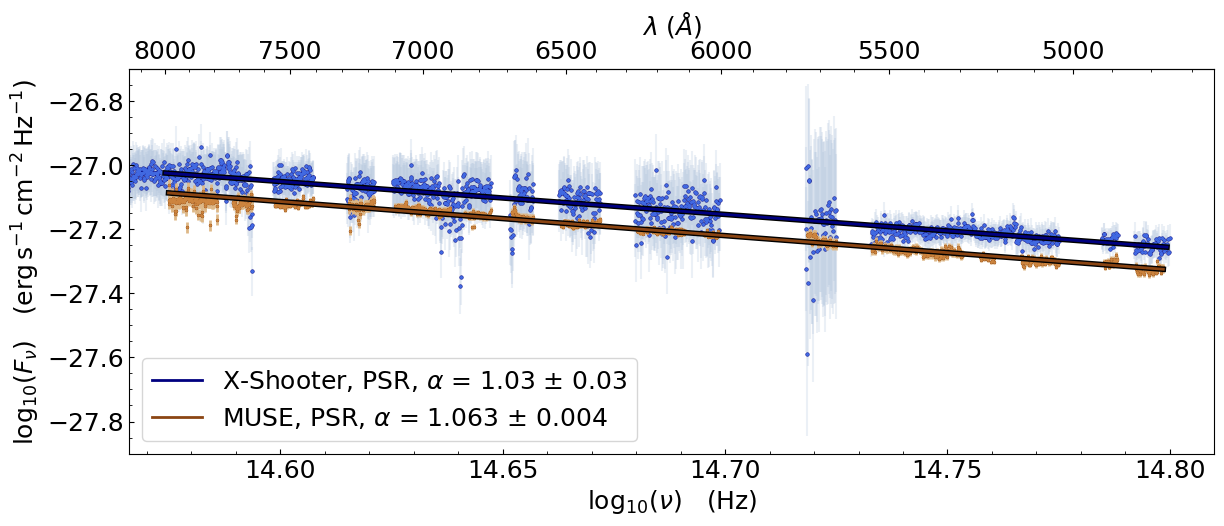}
\caption{Comparison of spectra from the pulsar region from \mbox{X-shooter} (blue points) and MUSE (brown points). The MUSE spectrum is extracted with a pseudo-slit resembling the \mbox{X-shooter} slit (constrained by the black lines in \autoref{fig:slit}). The PWN background has not been subtracted from these spectra. The two lines (black for \mbox{X-shooter} and brown for MUSE) represent PL fits to these spectra. Note that the \mbox{X-shooter} spectrum is binned in order to showcase where the largest amount of data points would lay. Non-binned \mbox{X-shooter} data completely overlaps the MUSE spectrum.}
\label{fig:appendix_comparison_pulsar_muse_vs_xs}
\end{figure*}

The spectra from this pulsar region and the results of the PL model fits (fit with the least-squares method for both instruments) are shown in \autoref{fig:appendix_comparison_pulsar_muse_vs_xs}. In this case the best-fit spectral indices are $\alpha=1.03\pm0.03$ for \mbox{X-shooter} and $\alpha=1.063\pm0.004$ for MUSE. We find that without the background subtraction, the pulsar region spectral indices are very similar to the corresponding PWN spectral indices. The reason for this is the significant nebula contribution in the pulsar region, which is $\gtrsim 60\%$ for both \mbox{X-shooter} and MUSE (see Appendix~\ref{app_sub:pulsar_systematics}). This effect is less prominent in the MUSE results, where the difference between the PWN and PSR is actually $>5\sigma$. We note that this is most likely due to the rather small fit parameter uncertainties, which might indicate that the flux uncertainties are underestimated.

The difference between \mbox{X-shooter} and MUSE pulsar spectral indices is $\sim0.03$, much smaller than in the PWN case above. We speculate that this is most likely serendipitous since the smaller extraction region for the pulsar should be more prone to exhibit differences due to systematic effects. We thus keep the estimate $0.1$ as the best measure of the systematic uncertainties in the spectral index measurements.

\begin{figure*}[t!]
\plotone{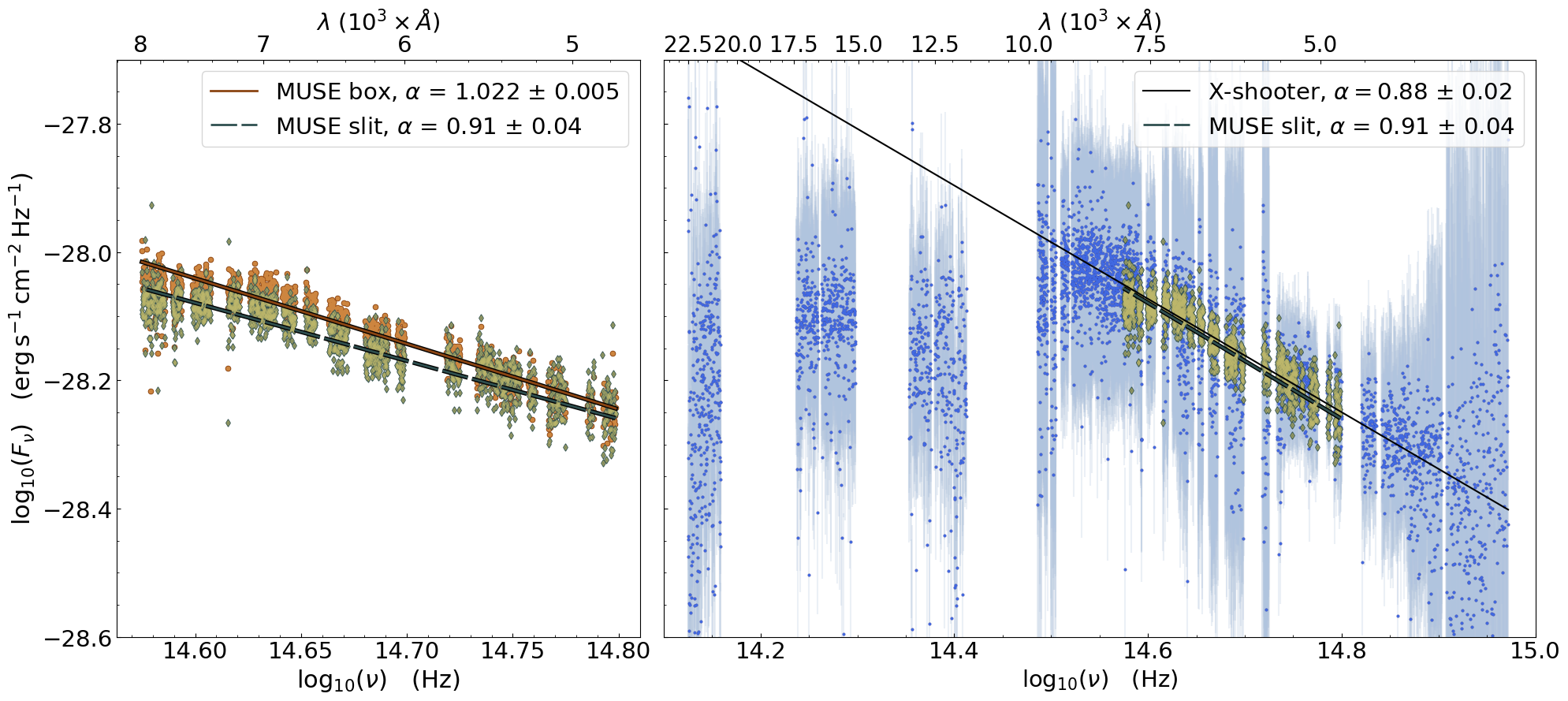}
\caption{\textit{Left}: Comparison of MUSE background-subtracted pulsar spectra extracted with a pseudo-slit resembling the \mbox{X-shooter} slit (\mbox{MUSE slit}, green diamonds) and the 3x3 spaxel aperture (\mbox{MUSE box}, golden circles). The two lines (dark green dashed line for \mbox{MUSE slit} and solid brown line for \mbox{MUSE box}) represent PL fits to these spectra. We have renormalized the flux through the \mbox{MUSE slit} to match the flux obtained with the \mbox{MUSE box}. \textit{Right}: Comparison of pulsar spectra from MUSE extracted with the pseudo-slit (\mbox{MUSE slit}, green diamonds) and from \mbox{X-shooter} (blue poits). Lines represent PL fits (dark green dashed line for MUSE and black solid line for \mbox{X-shooter}) to the MUSE wavelength range for both spectra.}
\label{fig:appendix_muse_pulsar_fit_continuum_1d_comparison_xs}
\end{figure*}

\subsection{Systematic Effects in Isolating the Pulsar Spectrum}
\label{app_sub:pulsar_systematics}

Here we investigate how the selected extraction regions affect the background-subtracted spectrum of the pulsar. The extraction regions for source and background in \mbox{X-shooter} are limited by the dimensions and placement of the slit, so it is important to determine how this affects the results. To this end, we start by comparing two different extraction regions in MUSE. In this section, we do not apply slit width corrections to the \mbox{X-shooter} spectrum, since we are studying a background-subtracted point source. A more advanced method for point-source slit width corrections would be to account for the flux losses caused by the \mbox{$\sim$\,0\farcs{8}--1\farcs{0}} seeing (that is wavelength dependent) with \mbox{1\farcs{2}--1\farcs{6}} slit widths following the method presented in \citet{Selsing2019}. We argue that this method is likely to produce a negligible effect in the case of PSR\,0540, since the emission is significantly dominated by the PWN. In addition, the background subtraction could be improved in general with time resolved data, see discussion in \citet{Sollerman2019}.

We use MUSE to probe how the pulsar spectrum changes when changing the extraction region by using two different apertures for extracting the pulsar spectrum. First, we use the slit-aperture (hereafter \mbox{MUSE slit}) defined in \ref{app_sub:agreement} above. Since the \mbox{X-shooter} slit width changes within the MUSE wavelength range, around \mbox{log-frequency}\,$\sim14.73$\,Hz, we multiply the flux values that lie below this log-frequency by the slit-width \mbox{ratio}\,$1\farcs{5}/1\farcs{6}$. We then compare this \mbox{MUSE slit} spectrum to the spectrum obtained with the $3\times3$ spaxel aperture (\mbox{MUSE box} hereafter) that is used throughout this work for the pulsar analysis.

We subtract the nebula contribution, which we continue to call background as in Section~\ref{subsec:pulsar}, from both spectra to remove the ${\sim 60\%}$ PWN contribution. For the \mbox{MUSE slit}, the background spectra were extracted from two 0\farcs{32} wide regions located 0\farcs{72} to the NE and 1\farcs{04} to the SW from the pulsar, respectively. In the NE, we left out 5 outlier spaxels. We note here that the PWN spectrum shows clear spatial variations and even changing a few spaxels can have a large effect on the resulting spectrum. In case of the \mbox{MUSE box}, a detailed description of the background subtraction can be found in Section~\ref{subsec:pulsar}. 

\begin{figure*}[t!]
\plotone{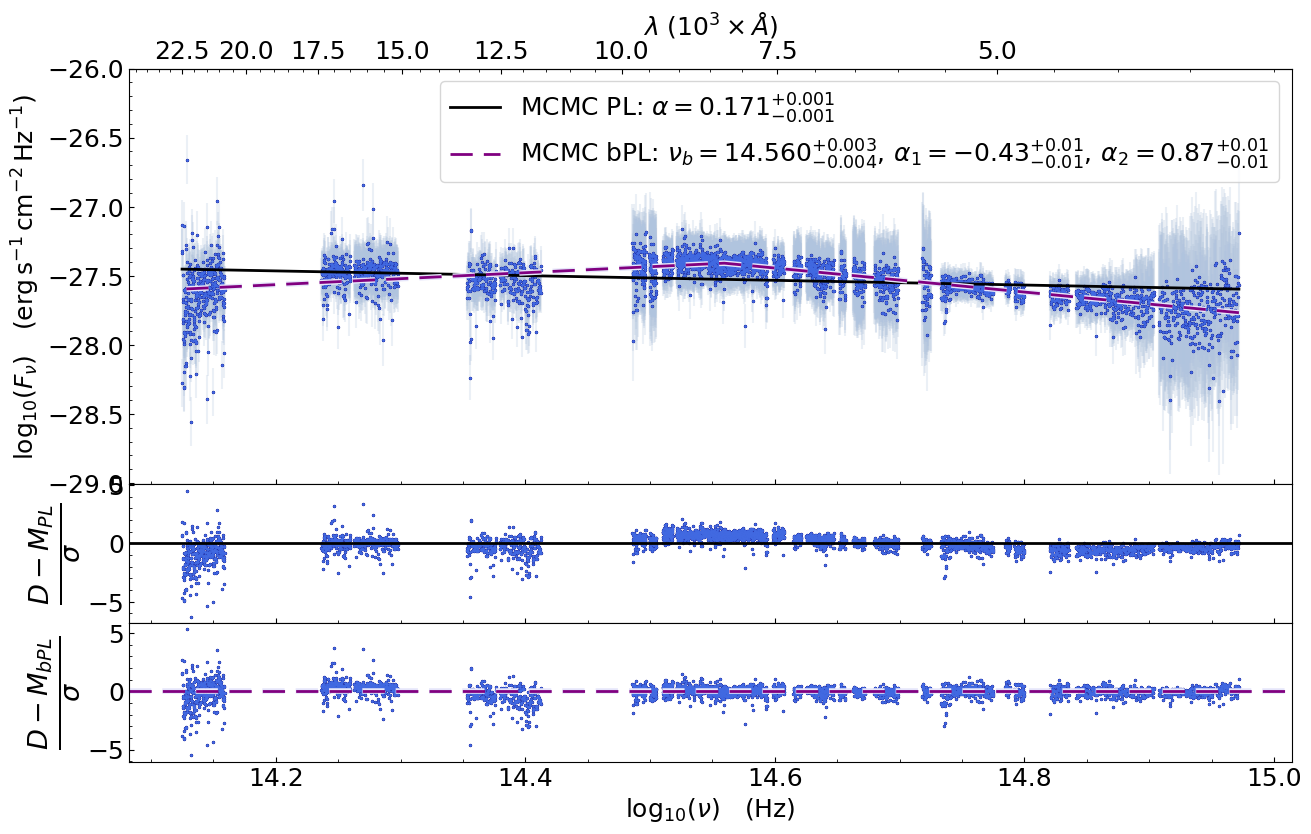}
\caption{\textit{Top:} \mbox{X-shooter}: background subtracted continuum spectrum from the pulsar, see \autoref{fig:slit} for the pulsar region. A PL model (black solid line) and a bPL model (violet dashed line) are fit to the entire UVB-NIR continuum emission.\textit{Middle:} Residuals of the PL fit. \textit{Bottom:} Residuals of the bPL fit.} 
\label{fig:appendix_xs_pulsar_continuum_fit}
\end{figure*}

The resulting background subtracted pulsar spectra for these two different extraction regions are shown in the left panel of \autoref{fig:appendix_muse_pulsar_fit_continuum_1d_comparison_xs}, where we also fit the spectra with a PL model. The spectral indices only agree to \mbox{within}\,$3\sigma$, the \mbox{MUSE slit} spectrum being harder ($\alpha=0.91\pm0.04$) than the \mbox{MUSE box} spectrum ($\alpha=1.022\pm0.005$). This is because the \mbox{MUSE slit} spectrum is extracted from a larger region and includes harder emission from the nearby nebula region in the NW and SE (see \autoref{fig:pixel_map}) compared to the \mbox{MUSE box} that covers a small region around the pulsar. This exercise showcases the high intrinsic spatial variability of the PWN and shows how it affects the background-subtracted pulsar spectra.

We also compare these results with the \mbox{X-shooter} background-subtracted pulsar spectrum. We use the same definitions for the \mbox{X-shooter} pulsar region and background as described above for the \mbox{MUSE slit}. The two spectra are plotted in the right panel of \autoref{fig:appendix_muse_pulsar_fit_continuum_1d_comparison_xs}. As expected, the \mbox{MUSE slit} spectrum overlaps with the \mbox{X-shooter} spectrum. In addition, we fit (with the least-squares method) the \mbox{X-shooter} background-subtracted pulsar spectrum (only MUSE wavelength range) with a PL model and obtain a slope of $\alpha=0.88\pm0.02$. Consequently, the \mbox{MUSE slit} and \mbox{X-shooter} background-subtracted pulsar spectral indices differ by $0.03$, which is  the same systematic offset found in Appendix~\ref{app_sub:agreement}. Furthermore, as can be seen in the right panel of \autoref{fig:appendix_muse_pulsar_fit_continuum_1d_comparison_xs}, the NIR range appears noisy and the connection between the NIR and VIS ranges is difficult to obtain due to the gaps in the NIR spectrum. This, together with the fact that the slit dimensions do not allow us to pick a sufficiently small extraction region, add uncertainty to the X-Shooter pulsar spectrum.

Keeping these uncertaintes in mind, we perform PL and bPL MCMC fits for the whole UVB-NIR range for completeness, since the single PL model fit for the MUSE range in the right panel of \autoref{fig:appendix_muse_pulsar_fit_continuum_1d_comparison_xs} significantly overestimates the emission in the NIR band. The best-fit PL model  (\autoref{fig:appendix_xs_pulsar_continuum_fit}) has spectral index ${\alpha_1=0.171^{+0.001}_{-0.001}}$. For comparison, the bPL model fit yields the break frequency ${\log_{10}\left(\nu_\mathrm{b}\right)=14.560^{+0.003}_{-0.004}}$ Hz, while the spectral indices are ${\alpha_1=-0.43^{+0.01}_{-0.01}}$ in the lower frequency range and ${\alpha_{2}=0.87^{+0.01}_{-0.01}}$ in the higher frequency range. As can be seen from the residuals, middle and bottom panel of \autoref{fig:appendix_muse_pulsar_fit_continuum_1d_comparison_xs}, the bPL model fit the spectrum better. These results confirm the break to a flatter slope at low frequencies found for MUSE (\autoref{fig:muse_pulsar_continuum_fit}), though the parameter values clearly differ.

\section{Extinction}
\label{app:extinction}

It is important to reliably estimate the extinction toward the source since this uncertainty significantly affects the derived spectral index. Earlier determinations of the color excess toward SNR\,0540 include $E(B-V) = 0.19$ mag \citep{Kirshner_1989} and $E(B-V) = 0.20$ mag \citep{Gordon_2003,Serafimovich_2004}, based on measurements of the Balmer line ratio near the SNR. To  assess the extinction based on our observations, we measure the flux ratios: Balmer lines $[H\alpha/H_\beta]_\mathrm{obs}$ by using the MUSE observations in Section \ref{subsec:hlines}, and the Fe-line ratio $[F_{1.257}/F_{1.644}]_\mathrm{obs}$, of the two emission lines [\ion{Fe}{2}] 1.257 $\mu$m and 1.644 $\mu$m using the \mbox{X-shooter} spectrum in Section \ref{subsec:ironlines}. 

The Balmer line ratio is measured for the surrounding ISM because for the SNR, the broad \mbox{H$\alpha$} line is blended with [\ion{N}{2}], while \mbox{H$\beta$} has low S/N. It is also not clear that the Balmer line ratio for case B recombination is applicable for the conditions in the ejecta. The Fe-line ratio, in contrast, is measured for the SNR. However, the value of the intrinsic Fe-line ratio is rather uncertain (as described below) and we therefore use the Balmer line ratio for the analysis.
 
\subsection{Color Excess from the Balmer Decrement}
\label{subsec:hlines}

The color excess $E(B-V)$ toward SNR 0540 can be determined by quantifying the Balmer decrement of the ISM H-clouds that overlap the central regions of \mbox{SNR 0540}. Because the intrinsic Balmer line ratio, $\left(\mathrm{H}\alpha/\mathrm{H}\beta\right)_\mathrm{int}$, stays approximately constant in standard gas clouds, the observed Balmer line ratio, $\left(\mathrm{H}\alpha/\mathrm{H}\beta\right)_\mathrm{obs}$, can be used to measure the decrement, which then leads to the estimation of the color excess \citep{Osterbrock-2006}:
\begin{equation}
E(B-V) = \frac{2.5}{k\left(\mathrm{H}\beta\right) -k\left(\mathrm{H}\alpha\right)}\log_{10}\left[\frac{\left(\mathrm{H}\alpha/\mathrm{H}\beta\right)_\mathrm{obs}}{\left(\mathrm{H}\alpha/\mathrm{H}\beta\right)_\mathrm{int}}\right],
\label{eq:color_excess_h}
\end{equation}
where $k\left(\mathrm{H}\beta\right)$ and $k\left(\mathrm{H}\alpha\right)$ are the values given by the reddening curve \citep{Cardelli_1989} at the correspoding wavelengths of the two emission lines. Assuming Case B recombination, temperature $T=10^4$~K, and density \mbox{$n_e = 10^2$ cm$^{-3}$} (for additional assumptions see \citealt{Junais-2023,Nebrin-2023} and references therein), yields to the intrinsic Balmer line ratio $\left(\mathrm{H}\alpha/\mathrm{H}\beta\right)_\mathrm{int}=2.86$ \citep{Osterbrock-2006}. As mentioned in Section \ref{sec:construction_of_the_continuum_spectra}, we use $R_V = 3.1$, which gives $k\left(\mathrm{H}\beta\right) = 2.53$ and $k\left(\mathrm{H}\alpha\right) = 3.61$.

\begin{figure*}[t!]
\plotone{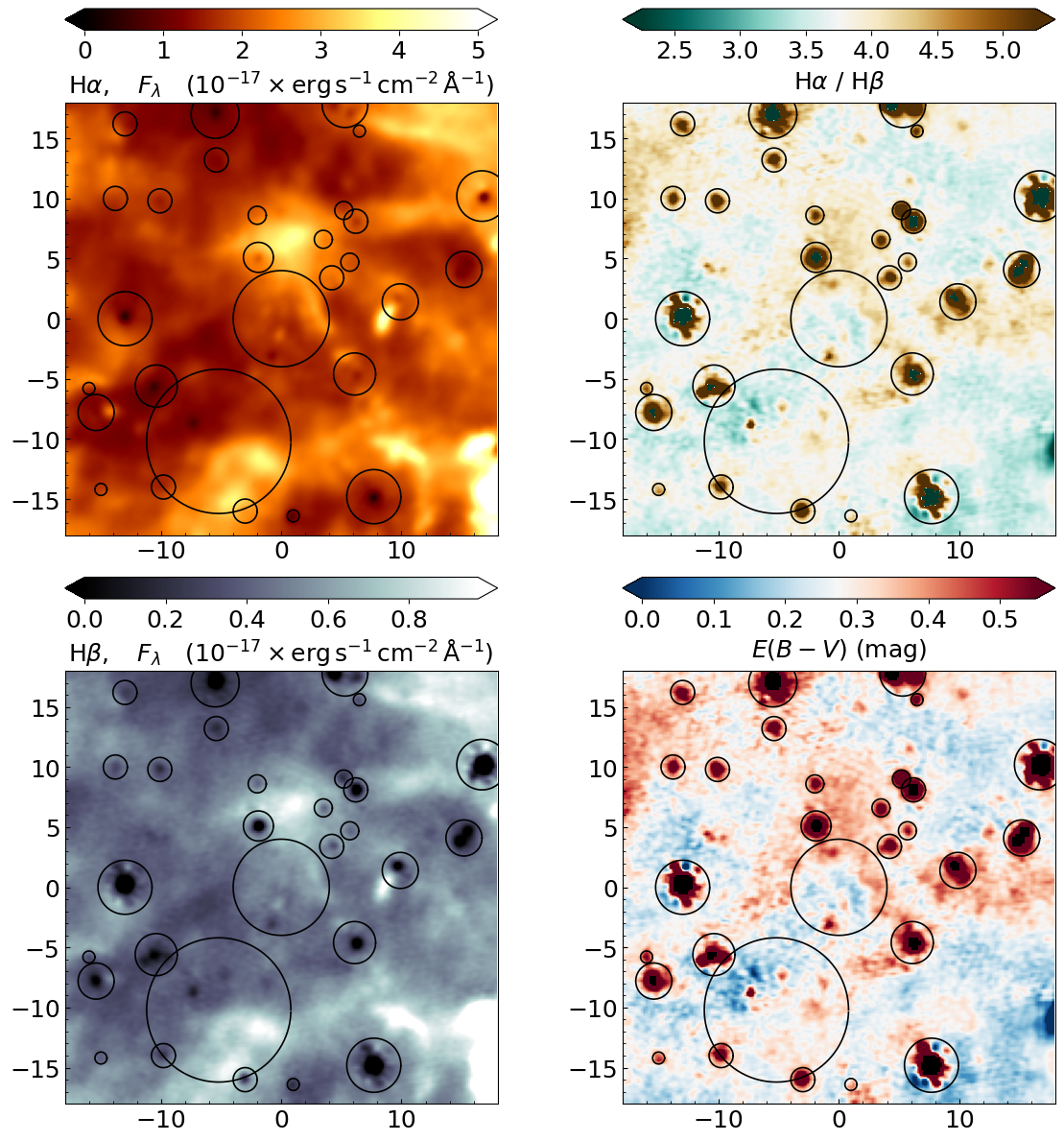}
\caption{Top and bottom left: MUSE images of the hydrogen lines H$\alpha$ and H$\beta$, respectively. Top right: Ratio of the two hydrogen lines. Bottom right: Color excess $E(B-V)$ computed from the hydrogen line ratio. The solid circles (radii \mbox{$0\farcs{5}$--$6\farcs{0}$}) define regions that were ignored in the color excess calculation due to the presence of stars (smaller circles) or emission from the SNR ejecta (the bigger circles, $4\farcs{0}$ radius at the center, and $6\farcs{0}$ radius at the SE, see \citealt{Larsson_21} for more details of the emission from the SNR ejecta). For the panels to the right, white color corresponds to the median value. The axes are in arcseconds centered at the pulsar in all panels.}
\label{fig:appendix_hydrogenlines}
\end{figure*}

The observed Balmer line ratio, $\left(\mathrm{H}\alpha/\mathrm{H}\beta\right)_\mathrm{obs}$, can be determined by studying the ISM hydrogen line fluxes observed by MUSE. We thus define a ${36\farcs{0}\times36\farcs{0}}$ region of interest centered at the pulsar to investigate the \mbox{H$\alpha$} and \mbox{H$\beta$} emission. In addition, we mask the innermost parts within $4\farcs{0}$ radius from the pulsar and a region with a $6\farcs{0}$ radius in the SE (\autoref{fig:appendix_hydrogenlines}) to avoid the hydrogen lines from the SN ejecta (L21). Additionally, we also mask regions that contain bright field stars.

We form 2D images of the hydrogen lines within this ${36\farcs{0}\times36\farcs{0}}$ region (\autoref{fig:appendix_hydrogenlines}); ${\sim\pm 200 \,\mathrm{km}\,\mathrm{s}^{-1}}$ around \mbox{H$\alpha$} and ${\sim\pm 270 \,\mathrm{km}\,\mathrm{s}^{-1}}$ around \mbox{H$\beta$} (interval width is $80\,\mathrm{\AA}$ for both lines). We subtract the continuum and background by determining background regions at $\sim -1400 \,\mathrm{km}\,\mathrm{s}^{-1}$ and $\sim +1500 \,\mathrm{km}\,\mathrm{s}^{-1}$ with widths $\sim 100 \,\mathrm{km}\,\mathrm{s}^{-1}$ for \mbox{H$\alpha$}, and at $\sim -1700 \,\mathrm{km}\,\mathrm{s}^{-1}$ and $\sim +1000 \,\mathrm{km}\,\mathrm{s}^{-1}$ with widths $\sim 135 \,\mathrm{km}\,\mathrm{s}^{-1}$ for \mbox{H$\beta$}. In this particular spatial region, the radial velocities of the Balmer lines do not vary significantly, so these simple cuts are sufficient to capture the whole extent of the lines. The median value of the Balmer line ratio within this region is then ${[\mathrm{H}\alpha/\mathrm{H}\beta]_\mathrm{obs} = 3.8 \pm 0.3}$, which allows us to finally compute the color excess ${E(B-V) = 0.27 \pm 0.07}$ with \eqref{eq:color_excess_h}. The uncertainties for both the observed Balmer line ratio and the color excess are one standard deviation. The color excess obtained here is higher than previous results.

\begin{figure*}[t!]
\epsscale{0.75}
\plotone{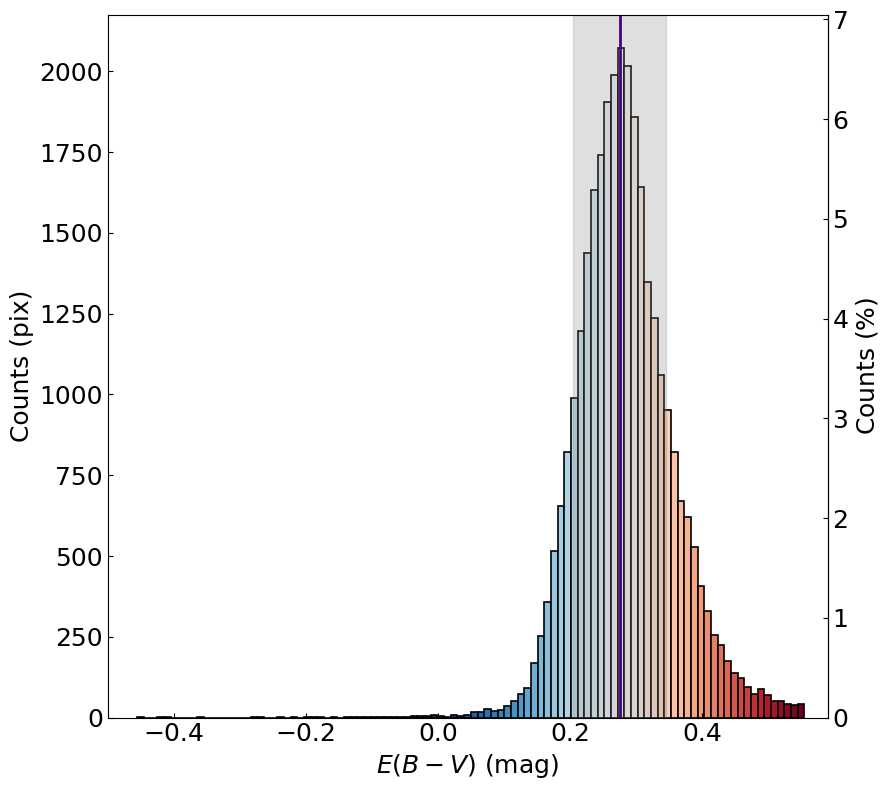}
\caption{Color excess values $E(B-V)$ computed pixel by pixel from the masked ${36\farcs{0}\times36\farcs{0}}$ region. The purple line indicates the median, and the grey region the standard deviation ${E(B-V) = 0.27 \pm 0.07}$. The color scale of the histogram is the same as in the bottom right panel of \autoref{fig:appendix_hydrogenlines}.}
\label{fig:appendix_histogram}
\end{figure*}

In \autoref{fig:appendix_histogram}, we further quantify the ${E(B-V)}$ distribution within the masked ${36\farcs{0}\times36\farcs{0}}$ region. This histogram reveals that there are no extreme colour excess values that would artificially force the estimated $E(B-V)$ higher. Additionally, to determine the color excess value, we use the median (not the mean) of the distribution, which further prevents the colour excess values located in the tails of the distribution affecting the result significantly. Despite the previous value ${E(B-V)}=0.2$ \citep{Serafimovich_2004} still being within one standard deviation from our higher value $0.27$, the histogram supports the usage of this higher value. Notable is that the previous results do not come with uncertainty estimations.

Finally, we address the main caveats affecting the color excess result. First, the H-regions considered here are assumed to be located near the remnant, which of course might not be the case due to projection effects. Second, some of the H-emission might also be coming from additional H-blobs originating from the SNR that are not masked in this analysis \citep{Lundqvist2022}. However, we note that these caveats concern also the previous color excess results. Since this work utilizes IFU data to determine the color excess, this result is likely statistically more significant and therefore we decide to use ${E(B-V) = 0.27 \pm 0.07}$ throughout this work.

\subsection{Color Excess from the Iron-line Ratio}
\label{subsec:ironlines}

\begin{figure*}[t!]
\epsscale{0.75}
\plotone{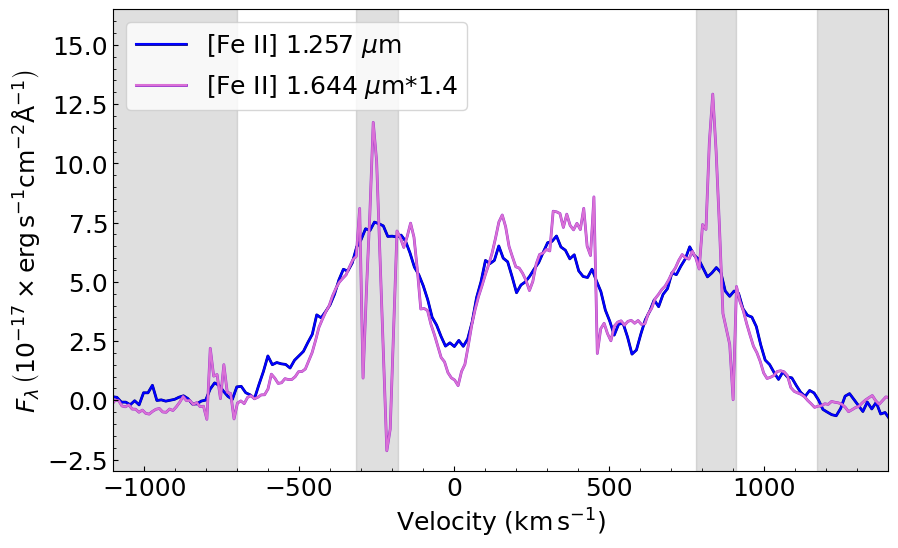}
\caption{Comparison of the two (continuum subtracted) iron lines [\ion{Fe}{2}] 1.257 $\mu$m and 1.644 $\mu$m in PWN\,0540. The grey areas indicate spectral ranges that are not included in the computation.}
\label{fig:appendix_ironlines}
\end{figure*}

We use data acquired from the nebula region (cyan lines in \autoref{fig:slit}), and subtract the continuum around the two iron lines. Since [\ion{Fe}{2}] 1.644 $\mu$m is noisy in regions from $-317$ to $-180$\,km\,$\mathrm{s}^{-1}$ and \mbox{780—910}\,km\,$\mathrm{s}^{-1}$ due to imperfect sky emission line corrections, we mask out these regions for both lines. To measure the flux ratio, we use fluxes within the range from \mbox{$-700$ to $1170$}{\,}km\,$\mathrm{s}^{-1}$ with the masks mentioned above. The flux uncertainties are estimated as standard deviations of the noise levels around the iron lines. The resulting lines are shown in \autoref{fig:appendix_ironlines}. We compute the flux ratio of ${[F_{1.257}/F_{1.644}]_\mathrm{obs} = 1.4\pm 0.2}$, where the uncertainty comes from the propagated flux uncertainties. We notice that the result significantly depends on the chosen region, which thus introduces a caveat for this computation.

By comparing the known intrinsic flux ratio $[F_{1.257}/F_{1.644}]_\mathrm{int}$ to the measured ratio obtained with our data, we can estimate the color excess $E(B-V)$ by \citep{Giannini2015,Koo2015}:
\begin{equation}
E(B-V) = \frac{1.086}{0.30}\ln\left(\frac{[F_{1.257}/F_{1.644}]_\mathrm{int}}{[F_{1.257}/F_{1.644}]_\mathrm{obs}}\right).
\end{equation}
The intrinsic iron line flux ratio $[F_{1.257}/F_{1.644}]_\mathrm{int}$, is still relatively uncertain, the reported values (both from theoretical and empirical studies) residing in the range 0.94—1.49 (\citealt{Lee_2015}, and references therein). Here we use both 0.94 and 1.49 to cover the full range of the possible values and get ${E(B-V)_{0.94} = -1.5 \pm 0.4}$\,mag and ${E(B-V)_{1.49} = 0.1 \pm 0.4}$\,mag for the color excess, respectively. Negative color excess values, in this case produced by the lower value of the intrinsic iron line flux ratio (0.94), are not expected. We are therefore able exclude the intrinsic iron line flux ratios in the lower range and estimate the color excess with ${E(B-V)_{1.49} = 0.1^{+0.4}_{-0.1}}$\,mag.

This result is consistent with the previously reported color excess values toward SNR\,0540 as well as the color excess value obtained above. The uncertainty in determining both the observed and the intrinsic iron line ratios prevent us from constraining this result further. Since the color excess result, ${E(B-V) = 0.27 \pm 0.07}$\,mag, is more constrained, we decide to use that value in this work.

\section{Pulsar Birth Period} 
\label{app:pulsar_age}

With a spin period of 50.77\,ms \citep{Marshall_16}, PSR\,0540 is one of the fastest spinning pulsars associated with a core collapse SN. Here we estimate the spin period at birth, using the improved kinematic age estimate ($t=1146\pm116$\,yr) from L21.

In general, pulsar period evolution can be approximated with an equation \citep{Shapiro_83, Vink_20} $\dot{\Omega} = - \kappa \Omega^n$, where $\Omega$ and $\dot{\Omega}$ are the pulsar's rotational frequency and its time derivative, respectively. Here $\kappa$ is a constant and $n$ the pulsar braking index. This differential equation can be solved for the pulsar birth period:

\begin{equation}
P_0 =  \frac{P}{\left(1 + t\cdot t^{-1}_{ch}\right)^{\frac{1}{n-1}}},
\label{eq:ST83}
\end{equation}

\noindent where $t$ is the time since the pulsar birth, $P$ the spin period at time $t$, $\dot{P}$ the spin-down rate, and $t_{ch}$ the characteristic age $t_{ch} \equiv \frac{1}{n-1}\frac{P}{\dot{P}}$. Here we assume that the characteristic age at birth equals the characteristic age at time $t$.

Instead of the canonical magnetic dipole model for the pulsar with $n=3$, we use the most recent braking index measurement for PSR 0540 by \citet{Ge_19} and set ${n=0.94\pm0.01}$. This gives a birth spin period of ${P_0=31.70\pm0.04}$\,ms.

There are several caveats in this estimate, which we briefly summarise here. First, PSR\,0540's braking index has been reported to vary significantly over the history from ${n\sim3.6}$ \citep{Middleditch1987} to  ${n\sim0.03}$ \citep{Marshall_16}. We, however, assume a constant braking index and use the most recently measured value for simplicity.

On the other hand, the measured braking indices, in general, might be affected by the choice of over-simplified models ($\kappa$ could evolve in time) or stochastic processes like timing noise (\citealt{Vargas-2023} and references therein). For example, \citet{Vargas-2023} demonstrate that accounting for this random timing noise in the pulsar braking index observations would render the anomalous values (${n\neq3}$) toward the canonical ${n\sim3}$ (although they do not discuss indices that initially fall in ${-2\lesssim |n|\lesssim 2}$). Therefore, we also compute the PSR 0540's birth spin period with the canonical breaking index ${n=3}$ and obtain a slightly larger birth period \mbox{$P_0=36.56$\,ms}.

It is thus clear that careful modelling of this pulsar and its properties is needed. Models that take into account events such as the 2011 spin-down rate change; possible time evolution of the braking index; and possible time evolution of the assumed constant $\kappa$ would be highly advantageous in understanding the evolution and properties of PSR 0540.

\bibliography{0540_continuum_accepted}{}
\bibliographystyle{aasjournal}



\end{document}